\documentclass[a4paper,11pt]{report}
\usepackage[T1]{fontenc}
\usepackage[utf8]{inputenc}
\usepackage{chngcntr}
\usepackage{palatino}
\usepackage{paralist}
\usepackage{microtype}
\usepackage{graphicx}
\usepackage{color}
\usepackage{subfig}
\usepackage[hyphens]{url}
\PassOptionsToPackage{hyphens}{url}
\usepackage{pdfpages}

\usepackage{amssymb}
\usepackage{amsmath}
\usepackage{array}

\usepackage[ocgcolorlinks, hyperindex]{hyperref}
\usepackage[xindy,toc,nonumberlist, acronym]{glossaries} 
\makeglossaries

\newglossaryentry{md}{name={molecular dynamics}, description={a method of computational chemistry that calculates the~movements of atoms caused by their interactions; this work deals with molecular dynamics with molecular mechanics model}}
\newglossaryentry{mm}{name={molecular mechanics}, description={a model of computational chemistry that considers atoms as undivided points of mass, it describes their interactions with empirical functions, it does not explicitly account for electrons or quantum effects}}
\newglossaryentry{fs}{name=fs, description={femtosecond, $10^{-15}$ s }}
\newglossaryentry{proteinfolding}{name=protein folding, description={a biochemical process in which a~protein folds, i.e. changes its 3D structure in fluent movement; commonly simulated by molecular dynamics}}
\newglossaryentry{conformation}{name=conformation, description={a distinct 3D structure of a~protein in a~local minimum of the~energy surface}}
\newglossaryentry{forcefield}{name=force field, description={a set of parameters and equations that empirically describe the~interactions of atoms}}
\newglossaryentry{energysurface}{name=energy surface, description={a function that describes the~energy of the~system with~respect to coordinates of all atoms}}
\newglossaryentry{samplingproblem}{name = sampling problem, description={the energy surface usually has many local minima split by energy barriers that are crossed only with some probability; it can take quite long for molecular dynamics simulation to cover all minima, i.e. to sample the~space of energy surface, hence the~sampling problem}}
\newglossaryentry{computationalchemistry}{name=computational chemistry, description={a field of chemistry that uses computational resources to model approximative representations of chemical systems}}
\newglossaryentry{solvent}{name=solvent, description={usually molecules of water and salt ions that surround the~molecule of interest in the~simulation}}
\newglossaryentry{molecule}{name= molecule of interest, description={a molecule of biological or chemical interest, e.g. a~protein, a~virus, a~na\-no\-tube}}
\newglossaryentry{explicit}{name=explicit model of solvent, description={the solvent is represented by explicit atoms of solvent molecules that take part in calculation in the~same way as atoms of the~molecule of interest}}
\newglossaryentry{implicit}{name=implicit model of solvent, description={the solvent is not explicitly represented by its molecules, its effects are included in the parameters of the system\comment{ \cite{Koehl2006}}}}
\newglossaryentry{strongscaling}{name=strong scaling, description={a function that maps increasing number of computational resources that solve the~problem of fixed size to the~wallclock time of the~computation; it shows how the~wallclock time (usually) decreases as the~resources grow but the~problem remains the~same}}
\newglossaryentry{weakscaling}{name=weak scaling, description={a function that maps increasing number of computational resources that solve the~problem of increasing size to the~wallclock time of the~computation; it shows how the~wallclock time ideally remains the~same as the~resources grow and the~problem grows}}
\newglossaryentry{wallclocktime}{name=wallclock time, description={the time duration of the~algorithm run as measured by wall clock}}
\newglossaryentry{computationtime}{name= computation time, description={the sum of time durations of the~algorithm run as measured by clocks of all processors that took part}}
\newglossaryentry{simulationtime}{name=simulation time, description={the time duration of simulated process; often number of steps $\times$ length of single timestep}}
\newglossaryentry{experiment}{name=experiment, description={an experiment in sense of traditional experimental chemistry}}
\newglossaryentry{insillicoexperiment}{name=in sillico experiment, description={an experiment done by computer simulation}}
\newglossaryentry{computerexperiment}{name= computer experiment, description={a series of runs of the algorithm that provide data for evaluation of algorithm properties}}
\newglossaryentry{exascalecomputing}{name=exascale computing, description={computing systems achieving the performance of exaFLOP/s, according to predictions exascale computing system will be built till 2020}}

\newacronym{mdacro}{MD}{molecular dynamics, in this work with molecular mechanics model}
\newacronym{mmacro}{MM}{molecular mechanics}
\newacronym{mmmd}{MM MD}{molecular dynamics with molecular mechanics model of interactions}
\newacronym{msm}{MSM}{multilevel summation method, a method for calculation of long-range interactions \cite{Hardy2006}}
\newacronym{fsacro}{fs}{femtosecond, $10^{-15}$ second}
\newacronym{exa}{exa}{$10^{18}$}
\newacronym{flop}{FLOP}{floating point operation}
\newacronym{flops}{FLOP/s}{floating point operations per second, a unit for measurement of performance of computing systems}
\newacronym{angstrom}{\AA}{angstrom, $10^{-10}$\,m, 0.1\,nm}

\definecolor{Orange}{rgb}{1,0.5,0}

\newcommand{\comment}[1]{}

\counterwithout{section}{chapter}
\counterwithin{figure}{section}

\begin{document}

\begin{titlepage}
\begin{center}
\textsc {Masaryk University}\\
\textsc{Faculty of Informatics}\\
\vspace*{30px}
\includegraphics[width=4.5cm]{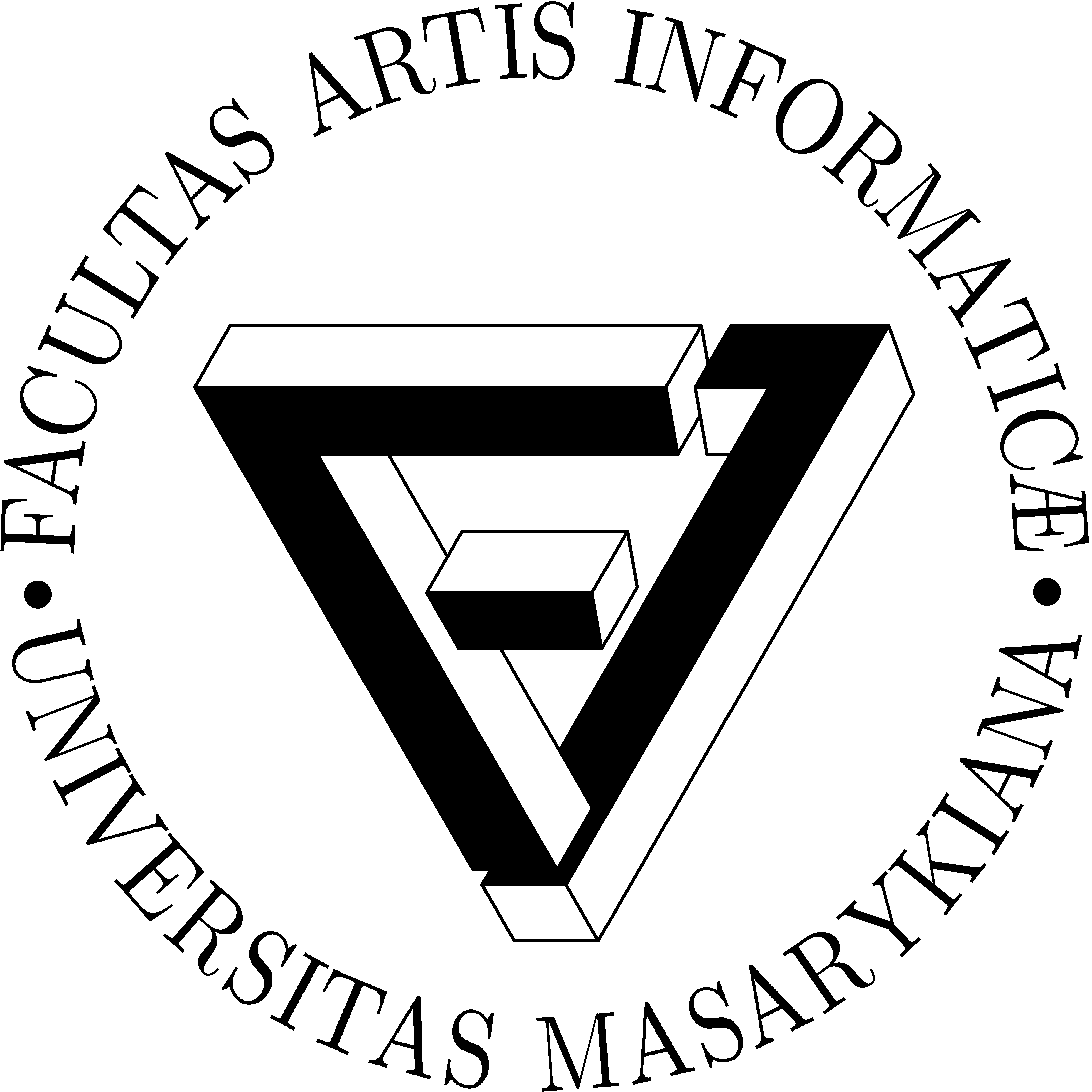}\\
\vspace*{20px}
{\huge \bfseries Large-Scale Molecular Dynamics Simulations\\[0.4cm]for Highly Parallel Infrastructures}\\[0.4cm]
\vspace*{70px}
\textsc{Ph.D. Thesis Proposal}\\
\vspace*{20px}
\textbf{\large Jana Paz\'urikov\'a}\\
\vspace*{80px}
\vfill
{\normalsize Brno, 2014}
\end{center}
\end{titlepage}

\pagenumbering{roman}
\setcounter{page}{2}

\phantom{.}
\newpage

\begin{tabular}{l}
\\
\end{tabular}

\vfill
\begin{tabular}{rl}
\bf Adviser: & {prof. RNDr. Luděk Matyska, CSc.} \\
&\\
\bf Adviser's signature: & \\
\cline{2-2}

\raisebox{4em}

\comment{\bf Consultant: & {RNDr. Ale\v{s} K\v{r}enek, Ph.D.} \\
&\\
\bf Consultant's signature: & \\
\cline{2-2}}
\end{tabular}
\newpage

\phantom{.}
\newpage

\noindent
Hereby I declare that this thesis proposal is my original authorial work, which I
have worked out by my own. All sources, references, and literature used or
excerpted during preparation of this work are properly cited and listed in
complete reference to the~due source.
\newpage

\phantom{.}
\newpage

\noindent
I would like to thank prof. Matyska and dr. Křenek for their supervision and consultation. The challenging task that led to this dissertation proposal has been suggested by dr. Vácha and dr. Kulhánek, I also greatly appreciate their advice on chemical aspect of work. Last but not least, I thank my colleagues from Sitola laboratory for their valuable comments.

\phantom{.}
\newpage

\tableofcontents
\newpage

\phantom{.}
\newpage

\pagenumbering{arabic}
\setcounter{page}{1}

\begin{section}{Introduction}
Computational chemistry allows researchers to experiment \textit{in sillico}: by running a~computer simulations of a~biological or chemical processes of interest. Computer models of the chemical processes offer higher resolution at the cost of decreased accuracy. Molecular dynamics (MD), a tool of computational chemistry, evaluates movements of particles caused by their interactions, it examines changes in time. Molecular mechanics (MM) models treat atoms as points with mass and charge and approximate their interactions with empirical functions. Molecular dynamics with molecular mechanics (MM MD) simulates N-body problem of atoms: it computes movements of atoms according to Newtonian physics and empirical descriptions of atomic interactions. Within each step the~computer evaluates the~forces exerted upon each atom caused by bond stretching, angle bending, torsion bending, van der Waals and electrostatic potentials and then moves the~atoms accordingly. In order to capture the~fastest 
oscillations occurring at atomic scale, the~vibrations of bonds containing the hydrogen, the~timestep of integration scheme is $\sim\,1$\,femtosecond ($10^{-15}$\,s). Common experiments simulate tens of thousands of atoms for hundreds of nanoseconds. \cite{Jensen2007, Lewars2010}

However, many interesting processes occur at longer timescales: tens of microseconds and more. Their simulations include tens, hundreds, thousands of billions of steps, each of them computationally demanding due to the~evaluation of electrostatic interactions. Therefore, MM MD simulations need high performance computing resources and approaches. 
Long wallclock time impedes the~research in areas that verify various proposed solutions to their research problems through MD simulations before using experiments, such as drug discovery or nanomaterial development.

Several different approaches deal with computational demands of MM MD. First, methods such as coarse-grained modelling \cite{Rudd1998}, discrete MD \cite{Proctor2011} and all algorithms for evaluation of electrostatic interactions \cite{Koehl2006} further simplify the~model and approximate its parts. Second, low level acceleration through specialized hardware \cite{shaw07} and GPU \cite{Stone2007,Anderson2008, Liu2008, Hardy2009} achieves high speed-up compared to computation on CPU, however, at increased cost. And last, computation on parallel and distributed infrastructures has been able to cut the~time to result thanks to spatial domain decomposition\cite{hess08, Case2005, phillips05, Christen2005, Plimpton2003, Arnold2013}. Unfortunately, with growing amount of computational resources, issues with scaling have arisen. 

Current parallel implementations of MM MD algorithms exhibit almost perfect weak scaling, i.e. they are able to simulate even large systems at reasonable wallclock time, if they are provided with enough computational power. However, the~fixed-size simulation will not run faster if provided with arbitrary number of computational resources, the~strong scalability hits the~wall. After reaching some critical limit in number of computing cores, adding more will not shorten the~time to result. The fine granularity of the~problem per core will result in high communication and synchronization overhead. The state-of-the-art MM MD algorithms with the~highest strong scaling can saturate up to half a~million cores \cite{Richards2009, Andoh2013}. With the~dawn of exascale computers, we want to shift the~level of parallelism further.

As the~spatial decomposition does not suffice, we propose rather uncommon approach: to calculate MM MD parallel-in-time.\comment{All current parallel and distributed implementations of methods compute parallel-in-space, i.e. they decompose the~spatial domain. Only the~Copernicus framework exploits quasi time-parallel computation when two processors calculate two possible paths from one conformation at the~same time. } Current research in molecular dynamics simulations is beginning to focus on time parallelism. Long simulations of protein folding have been conducted at highly distributed infrastructure through Copernicus and Folding@Home projects \cite{pronk11, Larson2002}. These frameworks exploit coarse-grained time parallelism by many parallel short simulations that explore the~conformation space of the~protein. Yu et al. \cite{Yu2006} use data from previous similar simulations to predict system states at future time points. Parallel-in-time methods, such as \cite{Nievergelt1964, Miranker1967, 
Vandewalle1994, Lions2001}, 
calculate the~results of a~time dependent differential equation in several successive time points simultaneously. This fine-grained form of time 
parallelism without \textit{a priori} knowledge is almost unknown in MM MD. Baffico et al. \cite{Baffico2002} wrote the~first, rather limited, publication on this topic in 2002. They concluded that parallel-in-time calculation ``can be very useful''. In 2013, Bulin wrote a~master thesis \cite{Bulin2013} where two parallel-in-time algorithms are compared in molecular dynamics simulations. \comment{He achieved the~speed-up 10 with the~parareal method and concluded that faster, yet still numerically stable coarse function $\mathcal{G}$ could lead to better scaling. We want to explore many possible choices of $\mathcal{G}$ function. } 

 One of the~parallel-in-time methods, the~parareal method, first approximates the~results ahead a~few timesteps with less accurate but cheap coarse function $\mathcal{G}$ and then iteratively corrects the~results in parallel with accurate but expensive fine function $\mathcal{F}$. The appropriate choice of coarse function determines the~convergence and speed-up of the~method. We have analyzed several combinations of fine and coarse methods and found two that promise high theoretical speed-up and reasonable convergence. We have designed the~parareal multilevel summation method with simple cutoff method or Wolf summation method as the~coarse functions and multilevel summation method as the~fine function. 

The aim of the~dissertation is to study and develop algorithms that would enable us to simulate large systems for long simulation times. We want to achieve it by incorporating the~time parallelism into the~calculation of long-range interactions in MM MD simulations. We will study and implement the~parareal multilevel summation method; evaluate its correspondence with results from experimental chemistry and compare the~accuracy, speed-up and scalability with the~best implementations of other methods. 

This thesis proposal continues with three more sections. First we will over\-view the~area of molecular dynamics, describe the~issues with the~calculation of long-range interactions, review several improvements that accelerate MD and introduce parallel-in-time computation. We will present state-of-the-art methods that are able to utilize large number of processors to speed up long simulations. In the~second chapter, we will state the~research questions and propose solutions. We will introduce the~novel parareal multilevel summation method\comment{, argument our choices for parareal method as scheme for time parallelism, multilevel summation method as fine function and both simple cutoff and Wolf summation method as coarse functions. We will }, analyze the~theoretical speed-up and convergence and suggest several directions of future work. In the~third chapter, we will state the~aim of the~work and schedule future progress.
\end{section}

\newpage
\begin{section}{State of the~Art}
\subsection{Molecular Dynamics Simulations in a~Nutshell}
\subsubsection{Introduction}  Molecular dynamics evaluates the movements of particles caused by their interactions. Many models describe the interactions: molecular mechanics, quantum mechanics, coarse-grained models and more. This thesis proposal deals with simulations of molecular dynamics with molecular mechanics model and the term \textit{molecular dynamics} (MD) will be used in this sense from now on if not specified otherwise. Through MD simulations, researchers can observe and experiment with a~model of mole\-cu\-les in order to understand chemical or biological processes and predict macroscopic properties by detailed knowledge of atomic movements caused by their interactions \cite{Jensen2007}. Bonded atoms interact due to bond stretching, angle bending and dihedral torsions. Non-bonded atoms interact due to van der Waals and Cou\-lomb interactions. Coulomb (also called electrostatic) interactions range to long distances, therefore they should be calculated between every pair of atoms. The evaluation 
of long-range interactions with current methods has $\mathcal{O}(N log N)$ or $\mathcal{O}(N)$ asymptotic complexity, where $N$ is the number of atoms. Still it remains the most demanding part of the calculation. Interactions can be described by the~potential function between the~particles and the~force caused by this potential changes the~positions of particles. The movement is governed by Newton's second law of motion commonly known as $\mathbf{F} = m\mathbf{a}$ where $\mathbf{F}$ is the~force, $m$ is the~mass of the~atom, $\mathbf{a}$ is the~acceleration of the~atom. Each step of the~simulation (that takes usually 2\,femtoseconds of the~simulation time), 
the~potential of interactions is calculated, Newton's second law of motion in form of the~partial differential equation is 
numerically solved and positions of atoms are updated. First computer simulations of molecular dynamics were published in 1950s \cite{Alder1959}; since then, many methods have been developed to accelerate the~calculation. Further approximations of electrostatic interactions have been designed and the~simulations are calculated on parallel and/or distributed infrastructure.

\subsubsection{Motivation} Thanks to molecular dynamics simulations, chemists and biologists can better understand how processes happen on atomic scale. That helps them for example:
\begin{enumerate}
 \item to simulate how a~protein folds or examine conditions under which it misfolds that can shed some light to the~cause of Alzheimer's disease or cancer \cite{Karplus2002, Snow2002, Lee2011};
 \item to see how enzyme interacts with a~substrate and how a~drug interferes with this interaction \cite{Boechi2013, Guimaraes2011, Schaefer2011};
 \item to see how designed nanomaterial interacts with other substances or reacts to the~external force that can foretell its properties \cite{Lau2012, Baweja2013, Tallury2010};
 \item to understand how phenomena occur and explain underlying physical reasons \cite{Ostmeyer2013, Zhao2013, Sotomayor2007}.
\end{enumerate}
MD simulations differ in the~size (number of atoms), simulation time (number of steps $\times$ length of single timestep). These characteristics and the~aim of simulations determine what further approximations or improvements can be done without major negative influence. 

Researchers may want to explore the~trajectory of each atom step-by-step. In that case, classical MD simulation provides the~data. However, various force fields (the set of parameters that determines the~calculation of atomic interactions) have been developed for various types of molecules---inorganic/organic, specialized/general \cite{Cornell1995, MacKerell2001,Halgren1996, Rappe1992}. Also, long simulations of small systems have different issues than short simulations of large systems when implemented for parallel or distributed infrastructure. 

Researchers may want to identify different conformations (3D structure of a~molecule in energy minimum) that the~molecule can get to from the~initial state and explore the~conformation space. Or they want to assess the~energy surface (the function that maps the~coordinates of atoms to the~energy of the~system). In these cases, they can accelerate the~computation by adding the~artificial, biased potential \cite{Laio2008, Torrie1977, Miron2004, Sugita1999}. The simulated system would not behave as it does in nature, however, unbiased properties can be reconstructed.

Molecular dynamics as a~tool of computational chemistry helps scientists in many fields and the~wallclock time of the~simulation crucially influences the phenomena they can study. 

\subsubsection{Simulation loop}
A general simulation algorithm first takes input data---types of atoms, the topology, partial charges $q_i$, positions $\mathbf{r}_i$ and velocities $\mathbf{v}_i$ and several other parameters for all particles in the~system---and then iteratively repeats the~following steps:
\begin{enumerate}
 \item {calculate the~potential and forces}
 \begin{equation}
 \begin{gathered}
  U_{all} = U_{bonded} + U_{non-bonded} = \\U_{bond} + U_{angle} + U_{torsion} + U_{Waals} + U_{Cou\-lomb}
  \end{gathered}
 \end{equation} 
 where $U_*$ is the~potential due to $*$;
 
 \begin{equation}
  \mathbf{F}_i =-\frac{\partial U_{all}}{\partial \mathbf{r}_i}
 \end{equation}
where $\mathbf{F}_i$ is the~force exerted on atom $i$;

 \item {move particles}
 
 \begin{equation}
 \label{eq:pde}
  \mathbf{F}_i =- \frac{\partial U_{all}}{\partial \mathbf{r}_i} = m_i \mathbf{a}_i = m_i \frac{ \mathbf{v}'_i}{dt} = m_i\frac{\mathbf{r}''_i}{dt^2};
 \end{equation}
 
 \item update time, optionally generate output. \cite{Jensen2007}
\end{enumerate}

The output of the~simulation includes the~trajectories of particles, forces and energy of the~system. Properties of the~system such as an average potential energy or the~viscosity of the~liquid are processed from output data by using further physical and chemical equations and applying statistical methods.

The differential equation (\ref{eq:pde}) is solved by numerical methods, most common are leap-frog \cite{Allen1989} and velocity Verlet \cite{Swope1982} integration schemes. Leap-frog scheme, a modification of Verlet method, starts with positions $\mathbf{r}$ at time $t$ and velocities $\mathbf{v}$ at time $t+\frac{1}{2}\Delta t$ where $\Delta t$ is the timestep. Then it updates the positions according to velocities, evaluates the potential and acquires acceleration in the next time point and updates the velocities \cite{Leach2001}:
\begin{equation}
 \begin{gathered}
  \mathbf{r}(t+\Delta t) = \mathbf{r}(t) +\Delta t \mathbf{v}(t+ \tfrac12 \Delta t) \\
  \mathbf{r}(t+\Delta t) \rightarrow U(\mathbf{r}(t+ \Delta t)) \rightarrow \mathbf{a} \\
  \mathbf{v}(t+ \tfrac32 \Delta t) = \mathbf{v}(t+ \tfrac12 \Delta t) + \mathbf{a}(t+ \Delta t)  \Delta t
 \end{gathered}
\end{equation}

Velocity Verlet starts with position and velocities at time $t$ and continues \cite{Leach2001}:
\begin{equation}
 \begin{gathered}
 \mathbf{v}\left(t + \tfrac12\Delta t\right) = \mathbf{v}(t) + \tfrac12 \mathbf{a}(t)\,\Delta t\ \\
 \mathbf{r}(t + \Delta t) = \mathbf{r}(t) + \mathbf{v}\left(t + \tfrac12\,\Delta t\right)\, \Delta t\ \\
 \mathbf{r}(t+\Delta t) \rightarrow U(\mathbf{r}(t+ \Delta t)) \rightarrow \mathbf{a} \\
 \mathbf{v}(t + \Delta t) = \mathbf{v}\left(t + \tfrac12\,\Delta t\right) + \tfrac12\,\mathbf{a}(t + \Delta t)\Delta t\ \\
 \end{gathered}
\end{equation}
More accurate methods, such as Runge-Kutta, would require the computationally demanding evaluation of the forces a few times within one step and therefore they can not be applied. 

The values for input data come from various sources. The structure of many proteins, viruses and other chemical compounds can be found in repositories such as \cite{Berman2003}. Partial charges of atoms are determined by the~force field. The initial velocities are randomly assigned by Maxwell-Boltzmann distribution corresponding to a low temperature. In a~short simulation, the~system is heated up to usually 300\,K by adding random kinetic energy to atoms. Often, similar short simulations follow to introduce the~solvent, minimize the~energy and stabilize the~pressure. After them, the~system is prepared for the~main simulation, relaxed and with stable pressure and temperature.

 \subsubsection{Issues and Limitations of Molecular Dynamics}
 \label{sec:mdissues}
\comment{ \begin{itemize}
  \item iterative solution to initial value problem
  \item high number of floating point operations in one step
  \item high number of steps to reach interesting timescales
  \item sampling problem
  \item force fields a~pevna topologia
  \item numerical instability, integration error
  \item communication necessary due to all-to-all interactions
 \end{itemize}}
Molecular dynamics has several issues, some caused by the~characteristics of simulated chemical processes, most of them caused by the~nature of the~method itself.

In models of computational chemistry, the~energy surface maps coordinates of atoms to the~energy. This function has many local minima that correspond to more or less stable states of the~system. In the~interesting processes the~system usually crosses the~energy barrier and transforms from one state to another. However, crossing the~barrier occurs with probability that exponentially relates to its~height, i.e. the higher barrier, the less probable crossing. Therefore, classical MD simulations sometimes have to simulate for long simulation time for crossing of the energy barrier to happen. As it can be rather difficult and lengthy to cover and sample whole energy surface, the~issue is called the~sampling problem \cite{Jensen2007}.   

Needed long simulation times directly lead to long wallclock times as MD solves the~initial value problem in a sequence of steps and, moreover, the~integration scheme has small timestep due to high oscillations of bonds that contain the hydrogen. The evaluation of the~potential between atoms in each step remains computationally demanding despite many approximations of long-range interactions. Moreover, evaluation of long-range interactions requires communication between all processors calculating spatially decomposed parts. Many decomposition techniques rather compute the~same values on two different processors than send a~message which stresses the~high temporal cost of communication \cite{Bowers2005}.

The relative error of integration scheme that affects the~forces achieves 1\% for common 2\,fs timestep\cite{Matthews2013}. This would cause unwanted changes in pressure and temperature. Therefore, MD simulation softwares regularly (every few steps) check the~temperature and pressure and alter the~atom velocities or the~volume to stabilize them. The simulations keep the~number of atoms, the~temperature and the~pressure or the~volume constant.

Despite errors due to rounding off, integration scheme, force field and approximations of long-range interactions, MD is considered to be accurate \cite{Skeel2005}. The~probability density of the~system states and global quantities of the~simulated system correspond with results acquired by experiments. Sometimes, the~simulated system blows up \cite{blowingup}. Blowing up refers to the~state with extremely large force that causes the~failure of the~integrator. The reasons the~system gets into such state include insufficient prior energy minimization, large timestep, inappropriate pressure or temperature control, unsuitable constraints and more. The simulation software needs to be carefully and properly configured as even the~small change in parameters influences the~simulation and can result in blowing up or in results that do not correspond with experimental results. 

Limitations of classical MD are based on its approximations and on the~characteristics of the~method. MD with molecular mechanics model works at atomic level, the~atom is an~undivided, mass and charge point. Electrons are not explicitly accounted, their distribution is usually represented by one scalar, the~partial charge. The topology of the~system (how atoms are bonded) stays the~same during the~whole simulation and no chemical reactions can happen\footnote{Although, force fields for some specialized cases of chemical reactions have been developed, e.g. \cite{VanDuin2001}.}. 

\subsubsection{Potentials}
\label{sec:potentials}
\comment{Movements of atoms in molecular dynamics are the~result of forces exerted on these atoms due to their interactions with each other. Bonded atoms interact due to bond stretching, angle bending and torsion angle bending. Non-bonded interactions between atoms in molecules may range to short distance, such as van der Waals interactions, or decay very slowly therefore have effect to long distances, such as Cou\-lomb (electrostatic) interactions. }All interactions are described by empirical functions that express the~potential of interactions with respect to the~distance between atoms (and other parameters). The form of these functions and the~values of their parameters together compose a~system called \emph{force field}. Various force fields have been developed, they differ in suitability for particular applicati\-on---some are more suitable for organic molecules, other focus on inorganic compounds. The most common are Amber \cite{Cornell1995}, CHARMM \cite{MacKerell2001}, MMFF94 \cite{Halgren1996} and 
UFF \cite{Rappe1992}.

\comment{Short-range potentials include all bonded interactions and van der Waals interactions. The bond between two atoms stretches and contracts, the~approximating equation is usually
\begin{equation}
  U_{bond} = \sum_{bonds\ i} k_i^{bond}(r_i-r_{0i})^2
\end{equation}
where 
$k_i$ determines the~``strength'' of the~bond: how difficult it is to stretch it. For example, this value tends to be higher in double bonds. {Check higher value, double bonds is the~correct term?}

The angle between three atoms widens and narrows, the~approximating equation is usually
\begin{equation}
 U_{angle} = \sum_{angles\ i} k_i^{angle}(\theta_i - \theta_{0i})^2
\end{equation}
where $k_i$ determines the~``strength'' of the~angle.

When four atoms connect sequentially (A-B-C-D), the~angle between two planes (A-B-C and B-C-D) called proper torsion angle changes. The approximating equation is usually
\begin{equation}
 U_{torsion} = \sum_{torsion\ i} k_i^{dihe}[1+ \cos(n_i\phi_i - \gamma_i)]
\end{equation}
where {dopisat vyznam $k_i$ a~$\phi$ a~$\gamma$ }.

When three atoms connect to the~one in the~center, the~improper torsion angles between them change. The approximating equation is usually
\begin{equation}
 U_{torsion} = \sum_{torsion\ i} k_i^{dihe}(\phi_i - \gamma_i)^2
\end{equation}
{dopisat vyznam $k_i$ a~$\phi$ a~$\gamma$ }
}

 All bonded interactions are short-ranged and they occur only between atoms connected by covalent bond. Bond stretching and angle bending are usually approximated by harmonic potential, torsion angles by cosine function. 
 Short-range non-bonded interactions, such as van der Waals interactions, London dispersion interactions and repulsions due to Pauli exclusion principle, decay very fast with increasing distance. Therefore, they are usually calculated only between atoms within short distance. Most common approximation is Lennard-Jones potential \cite{pumma}, expressed as  
 \begin{equation}
  U_{Waals} = \sum_{i} \sum_{j>i} 4\epsilon_{ij} \left[ \left( \frac{\sigma_{ij}}{\mathbf{r}_{ij}} \right)^{12} - \left( \frac{\sigma_{ij}}{\mathbf{r}_{ij}} \right)^6 \right]
 \end{equation}
where $\epsilon_{ij}$ is the~depth of a~potential well, $\sigma_{ij}$ is the~distance at which the~inter-particle potential is 0.

 As mentioned, short-range interactions of the~considered particle are calculated only with a~limited number of particles that are closer than preset distance. A few methods can construct a~list of neighboring particles for all atoms in $\mathcal{O}(N)$ time (for one atom in constant time) so that it is not necessary to check every pair of particles in the~system each step \cite{Plimpton1995}. First approach, called Verlet lists, includes in the~neighbor list of the considered particle all particles within extended cutoff distance $c_{extend} = c_{cutoff} + \delta$. Verlet list is rebuilt every few timesteps and $\delta$ is set so that no atom within $c_{cutoff}$ can move further than $c_{extend}$. Second approach, called link-cell method, bins all atoms to cells and for the~considered particle only particles in the~same cell or neighboring 26 cells are examined. The more efficient combination of these methods uses binned cells to create Verlet list.

 Short-range interactions, either bonded or non-bonded, are calculated very quickly as for each atom there is only small, upper-bound number of interacting atoms. They do not present a challenge from computational point of view but they crucially influence the movements of atoms. 
 
\subsection{Calculation of Long-Range Interactions}
The bottleneck of the~computation, electrostatic interactions ranging to long distances are calculated by Cou\-lomb's law between almost every pair of atoms (only atoms with bonded interactions are excluded) as
\begin{equation}
 \label{eq:coulomb}
 U_{Cou\-lomb} = \sum_i \sum_j \frac{q_i q_j}{|\mathbf{r}_i - \mathbf{r}_j|}.
\end{equation}
The research in methods of molecular dynamics focused intensively to approximate the~Cou\-lomb's law and reduce the~intrinsic quadratic complexity \cite{Koehl2006}. The simplest method---the cutoff method---calculates the~interactions by Cou\-lomb law but only for atoms that are within preset cutoff distance; it completely neglects the~interactions between atoms that are further apart. The method is rather inaccurate and it may introduce unphysical artifacts at the~edge  \cite{Loncharich1989, Schreiber1992, Beck2005}, however, it is fast and easily implementable. Smoothed cutoff method and Wolf summation method \cite{Wolf1999} deals with the~artifacts but the~accuracy remains rather low. 

All more sophisticated and more accurate methods interpolate the~char\-ges onto the~grid and then they calculate the~potential by
\begin{itemize}
 \item Fourier transform: Ewald sum \cite{Ewald1921}, Particle-Mesh Ewald \cite{darden93}, Smooth Particle-Mesh Ewald \cite{Essmann1995}, Particle-Particle Particle-Mesh method \cite{Hockney1988}, Gaussian split Ewald \cite{shan05};
 \item hierarchical division of the~space: Barnes-Hut method \cite{barneshut86}, Fast Multipole Method \cite{greengard87};
 \item multigrid methods \cite{Sagui2001, Skeel2002, Sutmann2005}.
\end{itemize}

\label{sec:methods}
Fourier transform methods are based on Ewald sum \cite{Ewald1921}. The electrostatic interactions between close atoms are calculated precisely by Cou\-lomb's law, the~long-range contribution is interpolated onto the~grid, calculated in Fourier space and then interpolated back from the~grid. Most software packages for MD simulation implement them, along with periodic boundary conditions they require. However, they do not scale well in number of processors used due to the~remaining many-to-many communication pattern. Most of the~methods based on Fourier transform have $\mathcal{O}(N\log_2 N)$ asymptotic complexity, however, they run equally fast or even faster than $\mathcal{O}(N)$ algorithms presented in the next paragraph. Implementation characteristics and code optimizations can efface the~difference, moreover, $\log_2N$ is below 30 even for the~largest systems. 

Methods based on the~hierarchical division of space stem from astronomical N-body simulations that resemble MD (instead of electrostatic potential they calculate gravitational interactions). Fast multipole method hierarchically divides the~simulation space into subcells and considers a~cluster of particles as one particle with combined charge from faraway point. Electrostatic interactions between particles in same or neighboring cells (on the~finest level) are computed directly, others are approximated by the~multipole expansion. It has $\mathcal{O}(N)$ complexity, however with a~large multiplicative constant. Moreover, the~implementation is rather difficult.

Multigrid methods, a~mathematical approach to solve partial differential equations \cite{Stuben1982, Briggs2000}, apply local process to multiple scales (grids) of the~problem. The algorithm calculates in V-scheme of grids: it goes from the finest grid to the coarsest grid and then back. First, it obtains an~initial approximation of results on the~finest grid. As the~error is smooth, the~corrections can be calculated on coarser grids by recursive relaxation (e.g.~Jacobi relaxation method) and restriction to coarser grids. The correction then prolongates from the coarsest grid to finer and finer grids. Their large multiplicative constant for $\mathcal{O}(N)$ complexity and iterative nature makes them difficult to implement for massively parallel resources.

Multilevel summation method \cite{Hardy2006} with $\mathcal{O}(N)$ complexity resembles multigrid methods. It calculates differently varying parts of potential on grids with different spacing. It exhibits valuable advantages: one-to-many communication pattern, rather simple parallel implementation and reasonable number of floating point operations. Therefore, we have selected it as the~fine function $\mathcal{F}$ for the~parareal method. Nevertheless, there are no characteristics of $\mathcal{F}$ function preventing other methods to take its place.


\subsection{Multilevel Summation Method}
 \label{sec:multilevel}
   Hardy in his dissertation \cite{Hardy2006} developed the~multilevel summation meth\-od (MSM) for calculation of long-range interactions with thorough mathematical background, performance assessments, accuracy analysis and implementation suggestions. The method divides the~calculation of the~potential onto multiple grids. To keep reasonable accuracy, more and more slowly varying parts of the~potential are calculated on coarser and coarser grids.

  

  Multilevel summation method hierarchically interpolates smoothed potential onto multiple grids and then sums it up \cite{Hardy2009}. The method therefore depends on two main functions---the smoothing function and the~interpolation function.
  
  With the~smoothing function $g_a(\mathbf{r}_i, \mathbf{r}_j)$, the~reciprocal distance $\frac{1}{|\mathbf{r}_{j}-\mathbf{r}_{i}|}$ can be rewritten so that it gives the~same result but its parts correspond to more and more slowly varying parts of potential.  
    \begin{equation}
  \label{eq:split}
  \begin{gathered}
   \frac{1}{|\mathbf{r}_{j} - \mathbf{r}_{i}|} = \left( \frac{1}{|\mathbf{r}_{j} - \mathbf{r}_{i}|} - g_a(\mathbf{r}_i, \mathbf{r}_j)\right) + g_a(\mathbf{r}_i, \mathbf{r}_j) \\
   g_a\left(\mathbf{r}_i, \mathbf{r}_j\right) = \frac{1}{a}\gamma \left( \frac{|\mathbf{r}_j-\mathbf{r}_i|}{a} \right) \\
   g_a\left(\mathbf{r}_i, \mathbf{r}_j\right) = \left( g_a(\mathbf{r}_i, \mathbf{r}_j) - g_{2a}(\mathbf{r}_i, \mathbf{r}_j) \right) + g_{2a}(\mathbf{r}_i, \mathbf{r}_j)
   \end{gathered}
  \end{equation}
where $a$ is the~cutoff distance and $\gamma$ is the~smoothing function.

The method has multiple grids in a~sequence and every grid considers doubled cutoff distance than the~grid before. So, the~grid in level $k$ has the~cutoff distance $2^ka$, where $a$ is the~cutoff on the~finest grid. Equations (\ref{eq:msmgrid}) reformulates equations (\ref{eq:split}), instead of subscript $a$ representing the~cutoff distance, they use superscript $k$ representing the~grid level. We will use the~superscript notation from now on.
  
\begin{equation}
\label{eq:msmgrid}
  \begin{gathered}
      \frac{1}{|\mathbf{r}_{j}-\mathbf{r}_{i} |} = (g^{*} + g^{0} + g^{1} + ... + g^{l-2} + g^{l-1})(\mathbf{r}_i, \mathbf{r}_j) \\
   g^*(\mathbf{r}_i, \mathbf{r}_j) = \frac{1}{|\mathbf{r}_{j}-\mathbf{r}_{i}|}-\frac{1}{a}\gamma\left(\frac{|\mathbf{r}_{j}-\mathbf{r}_{i}|}{a}\right) \\
   g^k(\mathbf{r}_i, \mathbf{r}_j) = \frac{1}{2^ka}\gamma\left(\frac{|\mathbf{r}_{j}-\mathbf{r}_{i}|}{2^ka}\right)-\frac{1}{2^{k+1}a}\gamma\left(\frac{|\mathbf{r}_{j}-\mathbf{r}_{i}|}{2^{k+1}a}\right) \text{ for } k=0..l-2\\
   g^{l-1}(\mathbf{r}_i, \mathbf{r}_j) = \frac{1}{2^{l-1}a} \gamma\left(\frac{|\mathbf{r}_{j}-\mathbf{r}_{i}|}{2^{l-1}a}\right) 
  \end{gathered}
\end{equation}
where $l$ is number of grids (grid levels $0..l-1$).
  
 $\gamma$ is an~unparameterized smoothing of the~function $\frac{1}{\rho}$ chosen so that the~first part of the~first equation (\ref{eq:split}), $\left( \frac{1}{|\mathbf{r}_{j} - \mathbf{r}_{i}|} - g_a(\mathbf{r}_i, \mathbf{r}_j)\right)$, vanishes after the~cutoff distance ($a$ for the~finest grid, $2^k a$ for grids of level $k$). In the~second part, $g_a(\mathbf{r}_i, \mathbf{r}_j)$, the~function $\gamma$ ensures partial derivatives and slow variation. Hardy usually sets it as Taylor smoothings, e.g. 
 \begin{equation}
  \gamma(\rho) = \left\{ \begin{array}{lr} \frac{15}{8}-\frac{5}{4}\rho^2 + \frac{3}{8}\rho^4, & \rho < 1,\\ 1/\rho, & \rho \geq 1 \end{array}\right. 
 \end{equation}
 
The interpolation function $\mathcal{I}$ makes it possible to do the~following approximation 
 \begin{equation}
 \label{eq:msmapprox}
  \begin{gathered}
    \frac{1}{|\mathbf{r}_i-\mathbf{r}_j|} = (g^* + g^0 + g^1 + ... + g^{l-2} + g^{l-1})(\mathbf{r}_i, \mathbf{r}_j) \\
    \approx (g^* + \mathcal{I}^0(g^0 + \mathcal{I}^1(g^1 + ... + \mathcal{I}^{l-2}(g^{l-2} + \mathcal{I}^{l-1}g^{l-1})...)))(\mathbf{r}_i, \mathbf{r}_j)
  \end{gathered}
 \end{equation}
 $\mathcal{I}$ interpolated $g$ to the~grid by 
  \begin{equation}
 \begin{gathered}
  \mathcal{I}^k g(\mathbf{r}_i, \mathbf{r}_j) = \sum_{\mu} \sum_{\nu} \phi_{\mu}^k(\mathbf{r}_i) g(\mathbf{r}_{\mu}^k, \mathbf{r}_{\nu}^k) \phi_{\nu}^k(\mathbf{r}_j)\\
  k=0,1...l-1
  \end{gathered}
 \end{equation}
 where $\phi_{\mu}^{k}$ is the~nodal basis function with local support (the function is non-zero only in the~close surrounding of the~grid point). For each grid with spacing $2^kh$, where $h$ is the spacing of the finest grid, and grid points $r_{\mu}^{k}$, nodal basis functions $\phi_{\mu}^k$ are defined as
 \begin{equation} \phi_{\mu}^k = \Phi\left(\frac{x-x_{\mu}^k}{2^kh}\right) \Phi\left(\frac{y-y_{\mu}^k}{2^kh}\right) \Phi\left(\frac{z-z_{\mu}^k}{2^kh}\right),\end{equation}  
 where $\Phi(\epsilon)$ can be for example the~cubic interpolating polynomial\footnote{Letters $\mu$ and $\nu$ represent grids or grid points, $k$ is the~grid level, $h$ is the~spacing of the~finest grid.}. 
 
 \comment{Now, we can insert some approximation by interpolating the~charges on coarser and coarser grids where their contribution will be added by $g$.

 \begin{equation}

  \begin{gathered} 
    \frac{1}{|\mathbf{r}_i-\mathbf{r}_j|} = (g^* + g^0 + g^1 + ... + g^{l-2} + g^{l-1})(\mathbf{r}_i, \mathbf{r}_j) \\
    \approx (g^* + \mathcal{I}^0(g^0 + \mathcal{I}^1(g^1 + ... + \mathcal{I}^{l-2}(g^{l-2} + \mathcal{I}^{l-1}g^{l-1})...)))(\mathbf{r}_i, \mathbf{r}_j)
  \end{gathered}
 \end{equation}
  where $g$ is defined in equation (\ref{eq:split}), $\mathcal{I}^k$ represents interpolation as described below.
 
 We have divided the~potential into several, more and more slowly varying parts. How do we put them on multiple grids and approximate the~solution?  as 
 
 Interpolation operator $\mathcal{I}^k$ interpolates each $g^k(\mathbf{r}_i, \mathbf{r}_j)$ to the~grid by 

}
 
 \comment{Then, we approximate $\frac{1}{|\mathbf{r}_i-\mathbf{r}_j|}$ as}

 With approximated reciprocal distance, the~potential is calculated as
 \begin{equation}
  U_i \approx \frac{1}{4\pi \epsilon_0} (u_i^{short} + u_i^{long})
 \end{equation}

 The following pseudoalgorithm represents the~equation written above. The short-range part within cutoff is $u_i^{short} = \sum_{j} g^*(\mathbf{r}_{i}, \mathbf{r}_{j})q_j$. Long-range part is recursively divided between two parts---one within the~cutoff and then calculated on the~current grid; and the~second representing even more slowly varying potential and then calculated on coarser grids. 
 The method for calculating the~long-range part that corresponds to evaluation of equation (\ref{eq:msmapprox}) goes as follows \cite{Hardy2009}:
 \begin{itemize}
  \item \textbf{anterpolation}---puts point charges onto the~grid $q_{\mu}^0 = \sum_{j}\phi_{\mu}^0(\mathbf{r}_j)q_{j}$;
  \item recursively for $k=0,1,...,l-2$
   \begin{itemize}
     \item \textbf{restriction}---approximates charges onto coarser grid \\$q_{\mu}^{k+1} = \sum_{\nu} \phi_{\mu}^{k+1}(\mathbf{r}_{\nu}^k)q_\nu^k$;
     \item \textbf{lattice cutoff}---calculates the~part of potential corresponding to the~grid $u_{\mu}^{k, cutoff} = \sum_{\nu}g^k(\mathbf{r}_{\mu}^k, \mathbf{r}_{\nu}^k) q_{\nu}^k$;
   \end{itemize}
  \item \textbf{top level}---calculates the~most slowly varying part of the~potential corresponding to the~coarsest grid $u_{\mu}^{l-1} = \sum_{\nu} g^{l-1}(\mathbf{r}_\mu^{l-1}, \mathbf{r}_{\nu}^{l-1})q_{\nu}^{l-1}$;
  \item \textbf{prolongation}---recursively backwards for $k=l-2, ..., 1, 0$ adds up the~parts of potential corresponding to the~grids $u_\mu^k = u_\mu^{k, cutoff} + \sum_{\nu} \phi_{\nu}^{k+1}(\mathbf{r}_{\mu}^k)u_{\nu}^{k+1}$;
  \item \textbf{interpolation}---puts grid potential off grid $u_i^{long} = \sum_\mu \phi_\mu^0(\mathbf{r}_i)u_\mu^0$.
 \end{itemize}

 The Figure \ref{fig:multilevel} shows a~specific example with three grids.

 \begin{figure}[h!]
 \begin{center}
 \caption{Example of multilevel summation algorithm.}
 \label{fig:multilevel}
  \includegraphics[width=\textwidth]{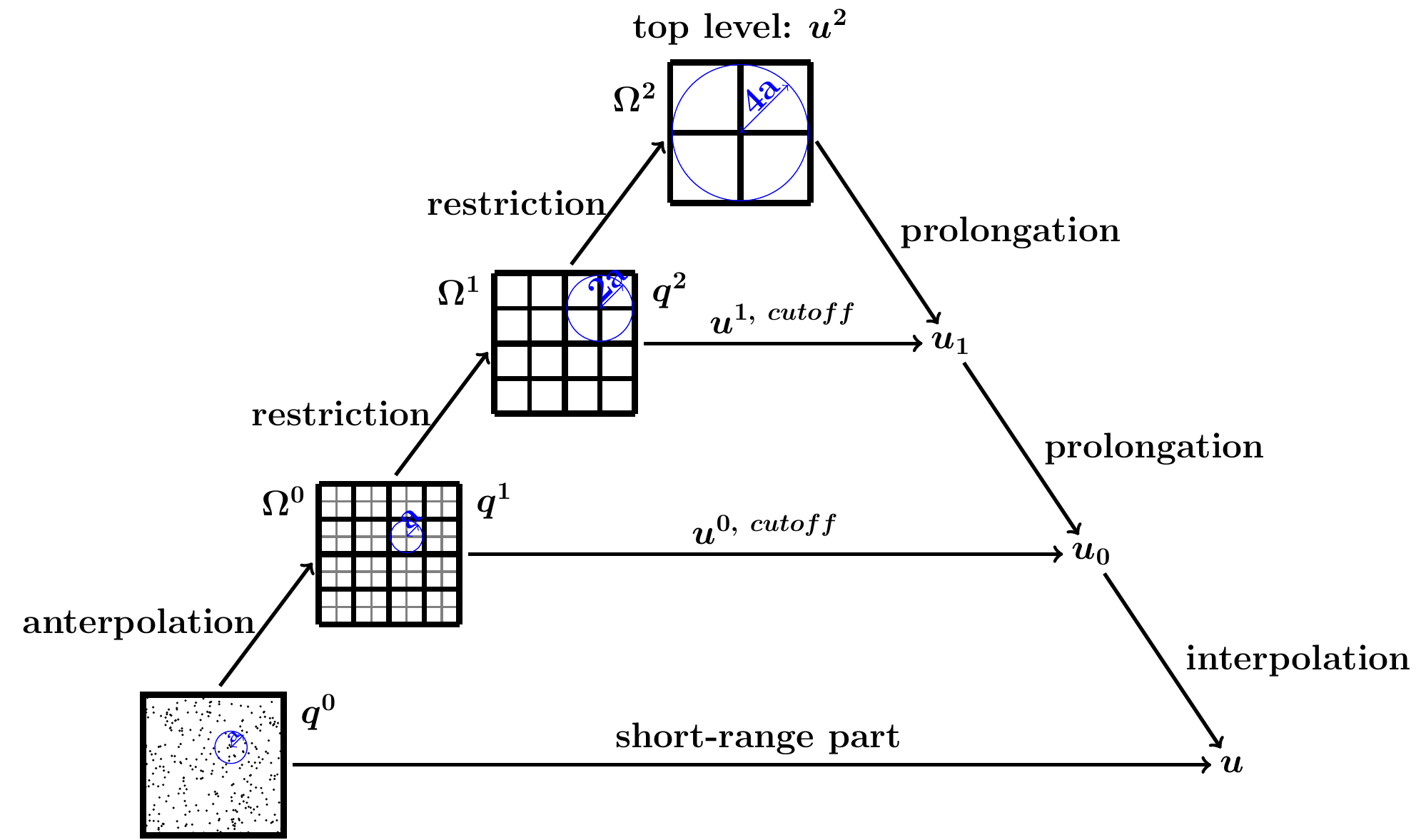}
  \end{center}
 \end{figure}

  Hardy precisely calculated the~number of floating point operations needed for the~evaluation of the~multilevel method \cite[ch.~2]{Hardy2006}. It is
  \begin{equation}
  \begin{gathered}
  \label{eq:flop}
   \left( \frac{4}{3} \pi m + \frac{32}{3}\pi + \frac{81}{2} \right) \left( \frac{a}{h^*} \right)^3 N +\\ \left( 6p^3 + 31p^2 + 36p + 17 \right) N + \left( \left( \frac{4a}{h} \right)^3 + 14(p+2)\right) \frac{8}{7} \left( \frac{h^*}{h} \right)^3 N
   \end{gathered}
  \end{equation}
 where
 \begin{itemize}
  \item $h^* = N^{-1/3}L$ is the~average distance between nearest neighbors 
  \item $h$ is the~spacing in the~finest grid 
  \item $a$  is the~cutoff distance 
  \item $N$ is the~number of atoms
  \item $L$ is the~length of the~box
  \item $\gamma$ is the~smoothing function polynomial of degree $2m$
  \item $\phi$ is the~basis function polynomial of degree $p$
 \end{itemize}
 
 For common parameters; $h^* = 1$\,\AA, $h=2$\,\AA, $a=12$\,\AA, $m=2$ (for $C^2$ Taylor smoothing function), $p=3$ (Hermite cubic interpolant) with relative error in force less than 1\% \cite[p.~78]{Hardy2006}; the~number of floating point operations reaches 
 \begin{equation}
  77.7a^3N+566N+\frac{73a^3}{h^6}N+\frac{80}{h^3}N \approx 136813N .
 \end{equation}
 
 \comment{In our simulation, $N=10^8$ and $L=500$\AA, therefore $h^* \approx 1$\AA. Usually \cite[in most evaluations]{Hardy2006}, $h=2$\AA~and the~cutoff $a$ ranges from $8$\AA~to $12$\AA. When choosing $\gamma$ and $\phi$ functions, we need to consider the~trade-off between the~accuracy and the~number of floating point operations. $C^2$ Taylor smoothing function ($m=2$) and cubic numerical Hermite interpolant ($p=3$) keep the~accuracy in reasonable boundaries (less than 1\% relative error in potential for $>8$\AA~cutoff \cite[p.~78]{Hardy2006}) while maintaining quite low cost. The number of floating point operations for these values is ($a=12$\AA)
 \begin{equation}
  77.7a^3N+566N+\frac{73a^3}{h^6}N+\frac{80}{h^3}N \approx 136813N \sim 10^{13}.
 \end{equation}}
 
 Multilevel summation method exhibits several advantages over common methods for calculation of electrostatic interactions \cite[ch.~1.2]{Hardy2006}. In comparison with fast multipole method (FMM), MSM calculates continuous forces and smooth potential. Therefore lower accuracy (e.g. lower order of interpolants) suffices for stable dynamics and conservation of energy. MSM, as multigrid and Fourier based methods, does not conserve linear momentum, FMM does. It is expected to fluctuate but not drift \cite{Skeel2002}, so it does not present an issue. When compared to Particle-Mesh Ewald method (PME), multilevel summation method scales better in parallel computation, uses multiple-time-step integration scheme more efficiently and communicates less \cite[ch.~7.4]{Hardy2006}. Apart from multigrid methods, MSM is not iterative therefore it scales better in parallel computation and does not require more than one global communication exchange. 

\subsection{Acceleration of Molecular Dynamics Simulations}
\comment{\begin{itemize}
 \item what are long simulations and why are they possible
 \item solutions - constraints
 \item multiple time step
 \item parallel and distributed computation - in the~next section
  \item GPU acceleration
 \item Anton - specialized hardware
 \item Copernicus, Folding@Home - highly distributed, Markov models
 \item losing atomic step-by-step scale - coarse models and discrete MD
 \item implicit solvent - greatly reducing the~number of atoms
 \item metadynamics

 \item conformational space exploration, protein folding - accelerated molecular dynamics, Copernicus
\end{itemize}}

Common MD experiments simulate hundreds of nanoseconds. However, many interesting biological and chemical processes occur at longer timesca\-les: tens of microseconds and more. Due to 2\,fs timestep, these long simulations require billions of steps, each computationally demanding. This subsection presents techniques that enable current MD software reach simulation times longer than hundreds of nanoseconds through model simplifications, computational volume reduction, hardware acceleration and parallel calculation.

The first type of methods further approximates and simplifies the~atomic step-by-step model of MD simulations with molecular mechanics. Higher abstraction leads to less demanding computation of each step or to less steps needed to simulate whole process of interest. 

Coarse models \cite{Rudd1998} describe interactions of beads that consist of a~few (usually tens of) atoms. The timestep of MD can be increased as the~fast movements of classical MM MD are hidden within the~bead. Thanks to that, molecular dynamics with coarse-grained models simulates up to 1\,ms processes. 

Discrete molecular dynamics \cite{Proctor2011} does not integrate Newton equations after fixed timestep, it assumes the~constant velocity of every atom until a~collision occurs. Then it recalculates velocities of collided atoms so that the~momentum is conserved. As the~interactions are calculated only within small number of atoms (only within those taking part in collision) and usually less frequent than every 2\,fs, the~simulation goes significantly faster compared to the~classical molecular dynamics. 

Conformation space mapping and energy surface approximation can complete their task much faster when the~simulation introduces biased, artificial potential to the~system to avoid staying in local energy minima. Metadynamics \cite{Laio2008}, umbrella sampling \cite{Torrie1977}, accelerated molecular dynamics \cite{Miron2004}, replica exchange \cite{Sugita1999} and similar address the~sampling problem with good results. 

Also, implicit models of solvent described below belong to this group.

The second type of methods eliminate the~parts of computation that usually give almost the~same values, such as interactions within the~water molecule.

Most of the~atoms in many simulated systems construct water molecules: up to 90\%. Systems need to be properly solvated due to two reasons. First, they give results better corresponding with reality. Without explicit water molecules, systems often include non-realistic physical artifacts due to changed electrostatics. Second, periodic boundary conditions, common in MD simulations, need at least 10\,\AA~layer of water to ensure that two replicas of the~molecule of interest do not ``see'' each other---their interactions are close to 0, thus negligible. In explicit modelling, water usually has a~rigid model that saves the~calculation of interactions within the~water molecule \cite{Horn2004, Rick2004, Toukan1985}. Implicit modelling does not explicitly consider solvent molecules, it includes its effects into the parameters of the system \cite{Vizcarra2005, Baker2005}.  That greatly reduces the~number of atoms, shortens the~time to result and enables quite fast multimillion-atom simulations
such as \cite{Schultz2009}. Hybrid models \cite{Lee2004} combine these two approaches: at 
the boundary of the~molecule of interest, explicit water molecules interact with it, further away there is the~reaction field. However, at the~interface between explicit and implicit representation, explicit water molecules do not behave in correspondence with reality. 

Thanks to constraints of bonds that contain the hydrogen, the~MD simulations have 2\,fs timestep. These bonds vibrate at frequency $10^{14}\ s^{-1}$ so the~timestep should be 1\,fs, constraints double that number. Most common algorithms are SHAKE \cite{Ryckaert1977}, LINCS \cite{Hess1997, Hess2008}, SETTLE \cite{Miyamoto1992}, RATTLE \cite{Andersen1983}. Force fields with united atom model (for example Amber) treat hydrogens differently than other atoms, they ``merge'' them with bonded atom (usually oxygen, carbon or nitrogen) and alter the~parameters for the~united atom \cite{Pearlman1995}. 

Multiple-time-step integration scheme takes into account that different interactions fluctuate over different timescales \cite{Allen1989, Grubmueller1998}. For example, the~potential due to bond stretching changes its value more often than the~van der Waals potential or electrostatic potential. Therefore electrostatic interactions can be evaluated less frequently than the~bonded interactions. 

The third type of approaches relies on hardware. One of the~longest simulations has been conducted on specialized hardware Anton in the~research of Shaw et al. \cite{shaw07} calculating ten microseconds of simulation time per day. The machine computation time has been offered for research calculations for free but naturally it is not capable to satisfy all needs of the~whole scientific community. Its cost prevents this concept to spread widely. Several MD simulation algorithms have been implemented on GPU \cite{Stone2007,Anderson2008, Liu2008, Hardy2009} with speed-up up from 10 to 100 (with multi-GPU even by three orders of magnitude) compared to optimized CPU code.

Parallel and distributed implementations of molecular dynamics code naturally speed up the~calculation; this topic is analyzed in the~next subsection \textit{Parallel Computation and Scalability}.

Presented methods and improvements make it possible to simulate even large systems (millions of atoms) for quite long simulation time (hundreds of nanoseconds) on rather available resources (thousands of cores). The acceleration leads to two goals: it is possible to simulate larger systems or it is possible to simulate the~same systems faster. If the~same simulation takes less wallclock time, longer simulation timescales are feasible and researchers from biology, chemistry or medicine can observe more and more interesting processes at atomic scale. 

\subsubsection{Parallel Computation and Scalability}
\label{sec:parallel}
\comment{\begin{itemize}
 \item why in parallel -> larger systems and faster (if faster, we can achieve longer timescales in feasible wallclock timescales)
  \item how they scale -> weak scalability, strong scalability
 \item how in parallel -> spatial decomposition \cite{Richards2013, Bowers2005}
 \item current state-of-the art algorithms -> urcite ten FMM z Juelichu, pozriet dalsie velke, ku nim konkretne data o skalovatelnostiach
  \item why is the~spatial parallelization not enough 
\end{itemize}}
The acceleration and parallel runs of MD algorithms follow two goals. First, to cut the~time to result. The wallclock time of the~computations crucially influences their usability and impact. The simulation that takes tens of minutes to complete on available resources can be repeated several times a~day and the~researcher can faster reveal mistakes or verify a~hypothesis---the simulations push the~research further. With short time to result, longer simulation times become feasible and previously unsimulated phenomena can be studied through MD simulations.

Second, to enable simulations with larger number of atoms. Fortunately, the~simulations with more atoms but also more computational resources tend to keep the~wallclock time similar. However, interesting chang\-es in larger systems take longer, which leads again to the~necessity of longer simulation times.

Types of scaling examined in MD algorithms correspond with the~two goals. First, the~strong scalability examines how the~wallclock time of fixed-size simulation changes (ideally reduces) as the~number of computational resources grows. For example, an~algorithm with an~ideal strong scaling would simulate the~same system million times faster on a~million-core supercomputer than on a~one-core computer. Naturally, every algorithm eventually hits the~strong scalability wall, when the~fine granularity of work reaches the~critical level and the~wallclock time increases due to the~communication and synchronization overhead. The research of parallel/distributed implementations of MD algorithms wants to shift the~wall further away and take advantage of massively parallel infrastructures.

Second, the~weak scalability examines how the~wallclock time changes (ideally remains the~same) as both the~size of the~problem (number of atoms) and the~number of computational resources grow, i.e. the~amount of work per processor remains the~same. For example, an~algorithm with an~ideal weak scaling would simulate 2000-atom system on two cores for the~same wallclock time as 1000-atom system on one core. Weak scalability makes it possible to simulate large system without major problems if enough computational power is provided. However, as mentioned above, large size of the~systems goes in hand with longer timescales necessary for relevant changes to happen.

All common MD algorithms have their parallel implementations with the~spatial decomposition in common software packages such as Gromacs \cite{hess08}, Amber \cite{Case2005}, NAMD \cite{phillips05}, GROMOS \cite{Christen2005}, LAMMPS \cite{Plimpton1995, Plimpton2003} or in libraries such as Scafacos \cite{Arnold2013, scafacos}. Bowers et al. in \cite{Bowers2005} overviews usual decomposition techniques. All parallel implementations exhibit almost perfect weak scaling, the~evaluations of strong scalability show the~wall commonly at tens of thousands of processors. The algorithms with the~highest strong scalability, to the~author's best knowledge, achieve the~saturation up to half a~million cores.

Andoh et al. \cite{Andoh2013} conducted a~simulation with $10^7$ atoms on over half a~million cores with simulation speed 35\,ns/day, one step was evaluated in 5\,ms. The long-range interactions were calculated by the~Fast Multipole Method. Richards et al. \cite{Richards2009} achieved almost 260\,TFLOP/s performance in the~simulation with $\sim10^9$ particles and showed almost ideal strong scaling up to 294000 cores. Copernicus \cite{pronk11} has simulated a~folding process of a~villin protein ($10^5$ atoms) through a~series of short simulations and Markov chain that connected them. They managed to reach remarkably fine granularity, 5000 cores simulated only 10000-atom system in wallclock time 3.5 day. Their results show the~potential of time parallelism incorporated into MD simulations.

The spatial decomposition does not suffice in three cases \cite{Srinivasan2005}. First, when the~simulated process has small number of atoms but it takes rather long. Second, when the~amount of work per processor gets low and the~fine granularity stops the~acceleration. And third, when the~communication is expensive as is common in distributed environments. All these three cases are becoming more and more common. Researchers want to simulate long processes due to their chemical or biological relevance but algorithms struggle to efficiently utilize increasing number of (sometimes distributed) computational resources at hand to make that possible.

Therefore, new approaches are needed to saturate more computational power and use it to shorten the~time to result. All state-of-the-art MD algorithms decompose only spatial domain and run parallel-in-space. We believe that the~parallelization along the~time domain would increase the~level of parallelism and shift the~strong scalability wall further. 
\comment{Naturally, all parallel implementations hit the~strong scalability wall: even with an~infinite number of computational resources, they would not be able to give results faster. With the~dawn of exascale computing and massively parallel infrastructures, }

\subsection{Parallel-in-Time Computation}
\comment{\begin{itemize}
 \item very short overview of parallel-in-time methods
 \item what can we gain and what we need to sacrifice, where is the~trade-off
\end{itemize}}
In parallel-in-time computation, different processors calculate results in different time points. Traditional, parallel-in-space computation decomposes the~spatial domain; different processors calculates results for different points in space. However, decomposition of the~temporal domain is not as easy due to the~sequential characteristic of time. Molecular dynamics sequentially solves the~initial value problem, so the~result from previous time points determines the~result in the~next time point. Several techniques are able to overcome this requirement, we introduce those concerning MD.

The project Copernicus \cite{pronk11} and Folding@Home \cite{Larson2002} use coarse-grained time parallelism. Many simulated processes include long but uninteresting metastable states of local minimum in which the~system gets stuck until it crosses the~energy barrier. For example, when a~protein is folding, it goes through different conformations, each of them is in the~local minimum of energy surface. The interesting part---how the~protein gets from one conformation to another---is rather quick compared to the~stay in metastable state. Projects build the~Markov model of different conformations that gradually explores the~conformation space by many short simulations. That allows them to use highly distributed framework and achieve remarkable strong scaling. Two different processors can simulate two possible paths from one conformation at the~same time, hence the~time parallelism. This approach is suitable for simulation of processes that consist of many metastable states separated by 
short transitions, e.g. protein folding or conformation space exploration.

Yu et al. in \cite{Yu2006} use data from prior related simulations to guide the~system and predict its changes. The simulation time is divided into intervals given to different processors. The prediction algorithm forecasts the~state of the~system at the~beginning of each interval according to previous simulations. That gives the~initial value for every processor and the~classical MD simulation through the~interval follows. At the~end of the~interval the~processor verifies that the~state corresponds with the~predicted state at the~beginning of the~next interval. If not so, the~data from the~mispredicted point in time are discarded, prediction algorithm adapts and the~simulation continues from the~latest correct state. The method has been evaluated on na\-no\-tubes pulled by the~external force, they achieved high strong scaling and speed-up. However, the~method relies heavily on reliability of the~prediction\comment{ that maps changes in prior base simulation with changes in 
current simulation}: the~presented high speed-up and 
strong scalability depend on no mispredictions during the~simulation. The prediction algorithm for nanotubes has been developed, no other suggestions for other MD applications (trajectory examination, conformation space exploration) have been mentioned.

Several mathematical methods have been developed that calculate par\-al\-lel-in-time in fine granularity: they compute results of a~time dependent differential equation in a~few (successive) time points simultaneously \cite{Gander2007}. The first such method by Nievergelt, 1964, \cite{Nievergelt1964} later became known as multiple shooting method. It divides the~time intervals to many subintervals, then solves initial value problem for each subinterval and forces continuity by Newton procedure. Time-parallel approach to iterative methods for solving partial differential equations with implicit integration schemes \cite{Deshpande1995} were followed by applying multigrid methods for acceleration \cite{Horton1992, Vandewalle1994, Horton1995}. In 2001, Lions, Maday and Turinici introduced \emph{parareal in time} method \cite{Lions2001}that has been extensively 
analyzed since then \cite{Gander2007, Maday2008, Maday2005, Aubanel2011}. 

Baffico et al. in \cite{Baffico2002} have done the~first MD simulation with the~parareal algorithm. The rather limited paper examined the~possible speed-ps and suitability of the~parareal scheme for classical MD and \textit{ab initio} MD that considers quantum mechanics as model for interactions. Shorter and longer timesteps were applied for the~fine and coarse function, respectively. They concluded that this approach is worth exploring and with many possible choices of fine and coarse functions, we are ``limited only by our imagination''.

Waisman and Fish \cite{Waisman2006} combined the~multigrid method \cite{Sagui2001, Skeel2002, Sutmann2005} and waveform relaxation method \cite{Lelarasmee1982} into space-time multilevel method with implicit integration scheme and speeded up MD simulation of a~polymer melt.

In 2013, Bulin published a~master thesis at Stockholm university \cite{Bulin2013} where he compared the~waveform relaxation method \cite{Lelarasmee1982, Vandewalle1994} and the~parareal method \cite{Lions2001}. Moreover, he suggested a~few improvements that could increase the~scalability of these methods. For the~parareal method, he proposed an~intuitive windowing (explained in the~following subsection) and applied multiple levels into the~parareal scheme. He achieved speed-up up to 10, the~number of iteration ranged from 4 up to 25 for different widths of the~window. Bulin concluded that the~waveform relaxation function is ``useless for this kind of problems'' due to slow convergence. For the~parareal algorithm, he stated that it is not suitable for large scale computing due to low speed-up. However, different coarse functions (faster, yet still reasonably accurate and numerically stable) could overcome this.

Despite rather disappointing results of parareal scheme in MD simulations, we want to research it further. Speck et al. in \cite{Speck2012} managed to simulate large gravitational N-body system through an~altered parareal method that nicely combined the~parallelization in spatial and temporal domain. As the~electrostatic and gravitational N-body problems have much in common, the~parareal scheme with a~more appropriate coarse function may succeed in MD simulation. 

\subsection{Parareal Method}

 The parareal method \cite{Lions2001, Bal2002, Maday2008} parallelizes the~time domain by approximating the~solution in time $t_{i+1}$ without accurate solution in time $t_i$ where $t_{i+1} > t_i$. When solving differential equation, we seek the~function of time $\mathbf{u}$ for initial condition $v$ in time $t>0$. 
%
  The exact solution is rarely known, more often we have the~precise enough approximation $\mathcal{F}_\tau(t, v)$ obtained by discretization with small timestep. $\mathcal{F}$ approximates $\{ \mathbf{u}(T_n) \}_n$ as
 \begin{equation}\{ \lambda_n = \mathcal{F}_{T_n-T_0}(T_0; v) = \mathcal{F}_{\Delta T_n}(T_{n-1}; \lambda_{n-1}) \}_n\end{equation} where $T_i$ are evenly-spaced time points. The sequential character of the~problem appears clearly: the~solution in time $T_n$ can be calculated only after the~solution in time $T_{n-1}$ is known.
 
 The parareal method proposes a~sequence $\{ \lambda_n^k \}_n$ that converges to $\{ \lambda_n \}_n$ rapidly as $k$, the number of iteration, increases and that can be built in parallel. It introduces the~second, coarse and cheap approximation $\mathcal{G}_\tau(t, v)$. The sequence $\{ \lambda_n^k \}_n$ is then defined recursively as
 \begin{equation}\label{eq:parareal}\lambda_{n+1}^{k+1} = \mathcal{G}_{\Delta T}(T_n; \lambda_n^{k+1}) + \mathcal{F}_{\Delta T}(T_n; \lambda_n^k)-\mathcal{G}_{\Delta T}(T_n; \lambda_n^k)\end{equation}
 The idea behind this sequence is to shift the~inherent sequential nature of calculation from $\mathcal{F}$ to $\mathcal{G}$. The equation (\ref{eq:parareal}) resembles predictor-corrector integration scheme \cite{Miranker1967, Gear1971}. Function $\mathcal{G}$ roughly assesses the~initial approximation of the~results. The difference between the~results from precise calculation and from coarse calculation on the~same data presents the~error that is included into calculation in the~next iteration. That gradually improves the~approximation of the~result. 
 
 In a long simulation with the~parareal scheme, the~width of ``computational window'' is determined by properties of the fine and the coarse function. Within the computational window, the results in successive time points are calculated parallel-in-time. As the~calculation proceeds (as explained in the~next paragraph), results from time points in the computational window converge and the~window shifts to the~right on the~time axis.
 
    \begin{figure}[h!]
  \caption{Computational flow of the~parareal method.}
 \label{fig:parareal}
 \begin{center}
  \includegraphics[width=\textwidth]{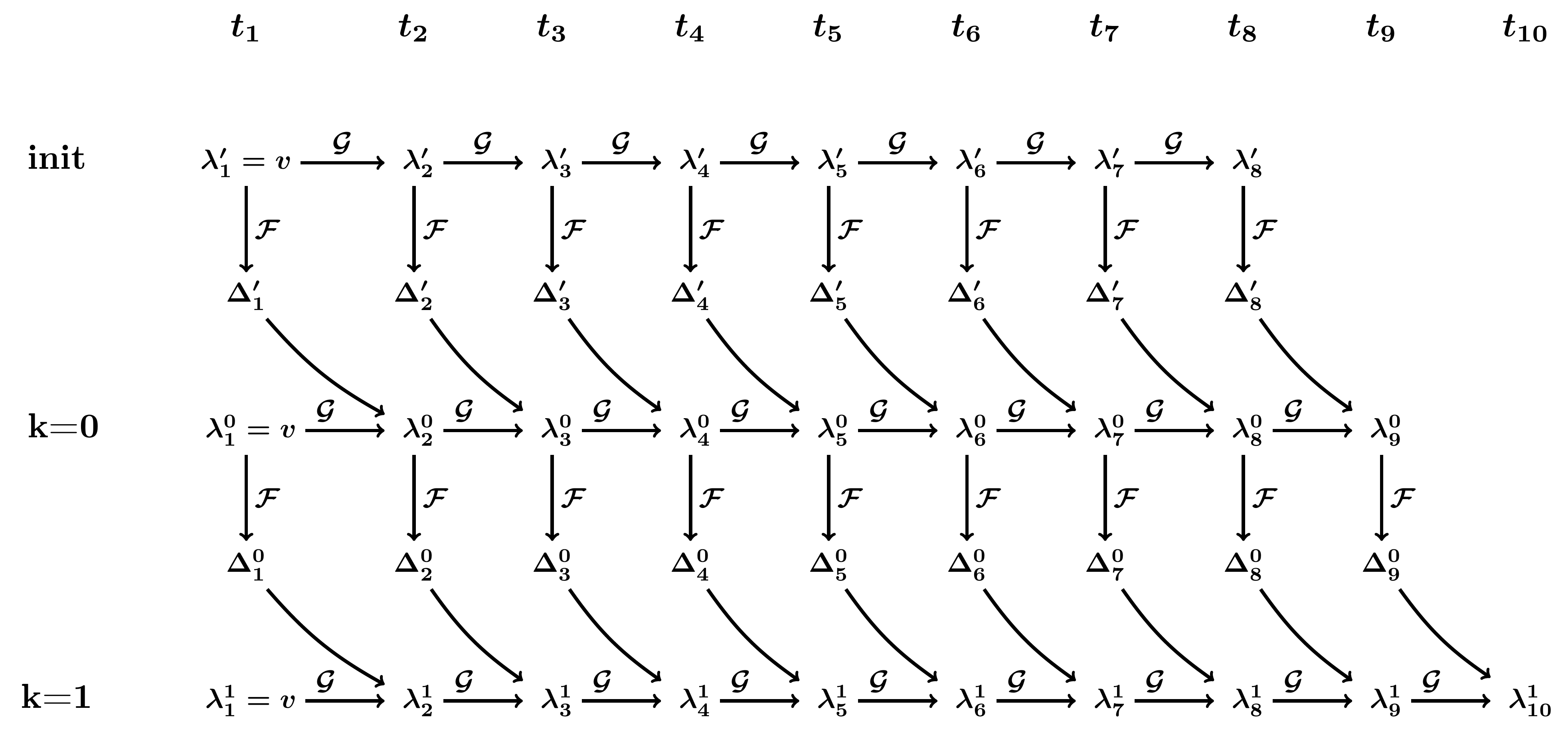}
  
\scriptsize
  $
\begin{array}{lllllll}
 & \lambda^1_1 = v &  & \lambda^1_2 \approx \lambda_2 \approx \mathcal{F}_{\Delta T}(T_1, v) & & \lambda^1_3 \approx \lambda_3 \approx \mathcal{F}_{\Delta T}(T_2, \lambda_2)&  \\
 \end{array}
$
$
   \begin{array}{lllllll}
  & &  \lambda^1_4 \approx \lambda_4 \approx \mathcal{F}_{\Delta T}(T_3, \lambda_3) & &\lambda^2_5 \approx \lambda_5 \approx \mathcal{F}_{\Delta T}(T_4, \lambda_4)  & ... &
 \end{array}
$
\normalsize
  \end{center}
 \end{figure}
  The Figure \ref{fig:parareal} shows an example of the~parareal method that calculates $\lambda_n$ for ten time points $n=1..10$ with the~initial condition $v$. The results converge after two iterations $k=0,1$. We apply further notation for clarity:
 \begin{itemize}
  \item $f_n^k = \mathcal{F}_{\Delta T}(T_{n-1}, \lambda_{n-1}^k)$
  \item $g_n^k = \mathcal{G}_{\Delta T}(T_{n-1}, \lambda_{n-1}^k)$
  \item $\Delta_n^k = f_n^k-g_n^k$
  \item $'$ superscript replaces $(-1)^{st}$ iteration---initialization
 \end{itemize}
 
 The arrows in the Figure \ref{fig:parareal} show the~computational flow---what needs to be computed in what order and with what dependencies. In \textit{init} row, the~calculation of $g'_n$ depends on already known $g'_{n-1}$. Downward arrows with $f_n^k$ and $\Delta_n^k$ represent parallel computation of the~error between the~precise approximation and the~coarse approximation done on the~same data\comment{\footnote{Inaccurate data does not mean inaccurate initial condition, we emphasize that the~\textit{coarse} approximation calculates results in time $t_n$ from results in $t_{n-1}$. This approximation gives us inaccurate data for calculation of result in the~next step.}}. In \textit{k} rows, $\lambda_n^k$ calculation depends on known $g_{n-1}^{k}$ and $\Delta_{n-1}^{k-1}$. After two iterations, we consider $\lambda_n^1$ very close to the~result (e.g. $\Delta_n^k<\epsilon$) of the~sequential computation of precise approximation based on accurate input data. The converged result in the last time point, $\
lambda_{10}^1$, will serve as the initial condition for calculation in the next computational window.
 
 The parareal algorithm multiplies the volume of computation, i.e. the number of floating point operations needed to get the results. Moreover, it requires several times more computational resources than the traditional, sequential-in-time integration schemes. However, if the number of iterations is smaller than number of time points, the method shortens time to result.  
 
 \subsection{Limitations of the~Parareal Method}
 The parareal method has its limitations: the~most important is the~relation between convergence and scaling. If the~method calculates in too many time points simultaneously, the~convergence begins to fail and the~number of iterations increases. And also too inaccurate coarse function with poor convergence and high number of necessary iterations assess well only results within a~few time points ahead.
 
 The maximum theoretical speed-up of the~method depends mainly on the~ratio of evaluation time of the~expensive $\mathcal{F}$ function and the~cheap $\mathcal{G}$ function, high ratio results in high theoretical speed-up. The high ratio is ensured by low evaluation time of $\mathcal{G}$ function, so cheaper functions are preferred. However, the~cost of the~coarse function coincides with accuracy. Cheap but too inaccurate functions cause lower convergence and higher number of iterations. 
 
 MD Simulations so far conducted with the~parareal method \cite{Baffico2002, Srinivasan2005, Nakano1993, Bulin2013, Baudron2013} achieved only low speed-up and efficiency or they exhibit issues with convergence. All of them used $\mathcal{G}$ function based on larger timestep. We believe that such function is unsuitable as it cause numerical instability in molecular dynamics and with more appropriate functions we can achieve better results.
 
 The convergence of the~parareal method has been extensively researched for both ordinary and partial differential equations, usually for Euler integration schemes \cite{Maday2005, Maday2008, Gander2007, Bal2005, Staff2003}. It has been proven that with any function $\mathcal{G}$, with no requirements on its quality, the parareal scheme will converge after at most $T-1$ iterations, where $T$ is the number of time points \cite{Gander2007}. Of course, in order to achieve some speedup, the number of iterations has to be much lower than number of time points, i.e. $K \ll T$. That depends on the stability of $\mathcal{G}$ function.  
 
 \comment{
 \begin{itemize}
  \item combination with spatial decomposition \cite{Maday2005}
  \item paragraph about the~disadvantages mentioned in \cite{Srinivasan2005} and \cite{Bulin2013} - speed-up limited to the~speed of $\mathcal{G}$, number of iterations, low efficiency with conducted experiments (mostly with larger timestep as coarse function)
    \item the~conditions of convergence general, proof in \cite{Maday2005}
 \end{itemize}
 }

 \subsection{Summary}
 Molecular dynamics offers biologists and chemists the~opportunity to observe processes at high resolution both in space and time. This work deals with molecular dynamics with molecular mechanics model of interactions. MM MD sequentially calculates the~forces exerted on all atoms caused by their interactions and then moves them according to Newtonian physics. The potentials described by empirical functions are evaluated in every step, they are computationally demanding due to the~long-range electrostatic interactions. Moreover, due to small timestep (usually 2\,fs), it takes large number of steps to reach interesting timescales (more than $10^6$ for short, 2\,ns simulations). Many approximations for Coulomb potential have been developed and their parallel implementations along with other improvements have made it possible to simulate millions of atoms for hundreds of nanoseconds. However, with increasing number of computational resources, issues with strong scalability have arisen. Only 
a~few of current algorithms are able to saturate over $10^5$ cores and their ability to run 
faster 
when 
provided with more processors hits the~limit. As all of them decompose only along the~spatial domain, we believe that computation parallel both in time and space would push the~level of parallelism further, thus increase the~strong scalability and harness even massively parallel computing infrastructures. First attempts to calculate MD simulations parallel-in-time have been made, however with little success. We want to study and develop the~parareal method and incorporate it in MD code. Apart from published approaches, we have selected different, more appropriate coarse functions that should provide high speed-up at relatively low cost and reasonable convergence.

\end{section}
\newpage
\begin{section}{Research Questions and Proposed Solutions}
\subsection{Research Questions}
\comment{\begin{itemize}
 \item explain the~request from dr. vacha
 \item how can we simulate such a~large system for such a~long time
 \item different method for MD: decrease the~number of FLOP - out of scope of computer science
 \item parallel and distributed implementation, moore's law, supercomputers, assessment of the~cores in the~largest supercomputer in ten years - how can we utilize them
 \item number of steps bigger problem than number of atoms, because spatial parallelism is good
 \item only spatial parallelism is not scalable for really large number of cores (or only a~few implementations)
 \item can we exploit some other form of parallelism - parallelism in time
 \item proposed solution presented below
\end{itemize}}

MD simulations have proven their usability for chemists and biologists. With faster algorithms and more computational resources, the~researchers naturally wish to simulate larger systems for longer simulation times. Parallel implementations of current methods manage simulations with higher number of atoms and short simulation time without major problems \footnote{The issues connected with large simulations cannot be underestimated. However, they are more of a~technical character as opposed to problems with long simulations times that come from the~fundamental concept of MD.}. However, processes concerning large systems usually take more time: it would be useless from biological point of view to simulate a~whole cell with $10^{14}$ atoms for a~nanosecond. Moreover, even many small systems take part in rather long processes of great biological or chemical interest. Current algorithms have almost reached their potential to cut the~time to result 
by adding more computational power that is available. 

A virus passing a~cell's membrane is an~example of rather long process (at least 50\,$\mu s$) with large number of atoms (hundreds of millions when properly solvated). Vácha et al.\comment{ and RNDr. Petr Kulhánek, Ph.D., CEITEC and the~National Centre for Biomolecular Research,  Dr. Vácha} has performed coarse-grained simulations in implicit solvent \cite{Vacha2011}, however they cannot answer some questions of chemical interest such as whether the~water molecules surround the~virus even when it is enveloped in the~membrane after the~passing. The full-atom simulation of properly solvated system would help them to observe the~process at close look. As computer scientists, we see the~challenge in \textit{how we can simulate such a~large system for such a~long simulation time}.

One possible way is to reduce the~number of floating point operations by changing the~MD algorithm, e.g. by inventing the~algorithm for calculation of long-range interactions with complexity $\mathcal{O}(\log N)$. However, that difficult task requires thorough background in chemistry, physics (to assess correctly what approximations can be made without major influence on accuracy), mathematics and computer science (to formulate N-body problem in such way that an~efficient implementation would be possible).

Another approach is to increase the~level of parallelization and utilize the~increasing number of computational resources at hand. Spatial decomposition that lays behind all current parallel implementations of MD code does not suffice for long simulation times as it scales only with the~size of the~system, not the~number of steps.\comment{ Spatial decomposition that lays behind all current parallel implementations of MD code, has almost perfect weak scaling---even large systems can be simulated without major problems if enough processors are added. However, parallelism in space does not suffice for long simulation times as it scales only with the~size of the~system, not the~number of steps.} The strong scalability of current algorithms hits the~wall at hundreds of thousands of processors. The largest supercomputer today has over three million cores\footnote{According to November 2013 list at http://www.top500.org, top 1 is Tianhe-2 at National Super Computer Center in Guagzhou, China, with 
3 120 000 cores and almost 55\,TFLOP/s peak performance.} and prognoses suggest that exascale supercomputer will be built till 2020 with $\sim 10^9$ 
cores. Therefore, we want to push the~strong scalability further to use more computational resources to shorten the~time to results. We strongly believe that parallelization in time would do 
that as the~scaling will then depend also on the~number of steps not only on the~number of atoms.

Time parallelism has been discovered for MD simulations. Coarse-grained parallelism in time has been developed in Copernicus and Folding@Home projects \cite{pronk11, Larson2002}. Even though their contribution to the~computational chemistry is undoubtful, they can not be applied for simulation of every biological process. The presence of metastable states and short length of interesting parts are two main conditions that not every process can satisfy. For example, when the~virus passes through the~cell's membrane, it is a~slow but continuous process researchers want to observe step-by-step. Yu et al. \cite{Yu2006} guide the~simulations of nanotube with the~previously acquired data from similar \textit{in sillico} experiments. However, it heavily relies on the~prediction algorithm. We consider to apply their approach in some way also to our method. Fine-grained time parallelism represented by mathematical parallel-in-time methods combined with MD have rather disappointed in previous research. Especially, 
the~parareal 
method has 
been analyzed, 
but the~coarse function $\mathcal{G}$ based on larger timestep led to unsatisfactory convergence and speed-up. 

We believe that with coarse functions based on more appropriate concepts, we can achieve better results. As the~fine function $\mathcal{F}$, we have chosen the~multilevel summation method. Therefore, we propose the~parareal MSM method, details follow in the~next subsection.

\subsection{Proposed Solutions}
\comment{\begin{itemize}
 \item combination of MSM and parareal
 \item what combinations we want to examine - simple cutoff, Wolf
 \item theoretical speed-up
 \item convergence issues
 \item adaptive control fo simulation with approximation of error
\end{itemize}}

 Our main research question focuses on how to simulate molecular dynamics of large systems for long simulation times. We want to increase the~strong scaling and the~level of parallelism by parallel-in-time algorithm to achieve computation parallel in both space and time. We propose a~novel method that combines the~multilevel summation method and parareal time integration method into parareal multilevel summation method: MSM will calculate parallel-in-space; the~parareal method will calculate parallel-in-time. The fine function $\mathcal{F}$ in parareal scheme will be classical MD with the~multilevel summation method for calculation of long-range interactions. The main characteristic of the~combination is the~choice of parareal method's coarse approximation $\mathcal{G}$. This function can be based on various concepts: 
 \begin{itemize}
  \item further simplification of the~model: discrete MD;\comment{, coarse-grained MD;}
  \item different parameters of MD algorithm: longer timestep;
  \item different parameters of the~method for evaluation of long-range interactions: coarser grid in MSM, shorter cutoff in MSM;
  \item less expensive method for evaluation of long-range interactions: simple cutoff method, Wolf summation method.
 \end{itemize} 
 
 We have considered several approaches and compared them to classical molecular dynamics with MSM. Moreover, we have examined two key aspects---the convergence and the~ability to run in parallel. 

    \paragraph{Discrete Molecular Dynamics} Coarse approximation $\mathcal{G}$ would run the~simulation using another concept---instead of integrating the~differential equations to capture the~movements of atoms we assume that they move with constant velocity unless a~collision occurs \cite{Proctor2011}. Despite its very good convergence, discrete molecular dynamics would not work as it produces completely different trajectories than classical molecular dynamics. The overall properties of the~system are similar but the~atoms move differently. As we are particularly interested how exactly atoms move (for example, as the~virus passes the~membrane), this approach does not suit us.
    
    \paragraph{Longer Time Step} Coarse approximation $\mathcal{G}$ would have longer timestep in integration scheme, instead of $2$\,fs it would be e.g.~$10$\,fs. The problem would appear in the~beginning of the~computation, before $k=0$, when we need to calculate $\lambda'_{n+1} = \mathcal{G}(t_{n+1}, t_n, \lambda'_n)$ where $\lambda'_1 = v$. The coarse approximation $\mathcal{G}$ therefore needs to give at least a~little reasonable results for every $t_n$ knowing only initial condition $v$ and its own result for $t_{n-1}$. Unfortunately, MD simulations with timestep larger or equal 5\,fs give unusable results after just a~few steps (in our own experiment with retinol and timestep 10\,fs the~system blew up after two steps). Half-converged results, e.g.~$\lambda_n^3$, instead of $\lambda'_n$, could save the~convergence but we would have to shorten the~computational window. That would drastically reduce the~ability to run in parallel. A few simulations of the~parareal scheme with longer 
timestep coarse function have been conducted \cite{Baffico2002, Bulin2013, Baudron2013}, with relatively modest speedup for larger simulations (lower than 10) and issues with convergence (up to 25 iterations). 
   
    \paragraph{Coarse Grid in MSM} Coarse approximation $\mathcal{G}$ would run molecular dynamics with multilevel summation method for evaluation of long-range interactions but the~finest grid would have spacing e.g.~four times larger than in the~fine approximation $\mathcal{F}$. It would be easily implementable, however, the~complexity assessment suggests it would not offer high speed-up as the~number of floating point operations depends mainly on the~cutoff distance $a$, not on the~finest grid spacing $h$.
    
    \paragraph{Shorter Cutoff Distance in MSM} Coarse approximation $\mathcal{G}$ would run molecular dynamics with MSM but the~interactions on the~current grid will be calculated within shorter cutoff range\footnote{Cutoff distance in MSM is the~distance within which the~part of potential is calculated on current grid.} than in the~fine approximation $\mathcal{F}$ (e.g.~6\,\AA~vs 12\,\AA~for the~finest grid). Larger parts of the~potential would be calculated on coarser grids, eventually large part of the~potential would be calculated on the~top level where it is cheaper than on finer grids. In evaluations \cite[ch.~3-5]{Hardy2006}, Hardy found the~relative error of forces to be in range from 1\% to 5\% with cutoff beginning at 8\,\AA. Our analysis showed that the~maximal theoretical speed-up is 7. As simple cutoff method or Wolf summation offer much larger theoretical speed-ups, we abandon this possibility. 
    
    
     \paragraph{Cutoff Method} Coarse approximation $\mathcal{G}$ would calculate long-range interactions by the~simplest method available---sum up the interactions with Cou\-lomb equation between atoms within constant cutoff, usually set to 12\,\AA. The accuracy does not reach the~level of elaborate methods but the~precise $\mathcal{F}$ function should correct the~errors. The rather good convergence of the~method should enable large width of the~computational window that determines the~ability to run in parallel, and low number of iterations. The smoothed cutoff method that gradually vanishes the~interactions between atoms distant further than cutoff distance should raise the~accuracy little up in case that simple cutoff method fails. The method is computationally cheap, so the~ratio between evaluation time for $\mathcal{F}$ and $\mathcal{G}$ that determines the~speed-up is quite high. 
    
    \paragraph{Wolf Summation Method} Coarse approximation would use Wolf summation method \cite{Wolf1999} to calculate electrostatics. We assume that with little additional computational cost (compared to the~cutoff method), we can gain faster convergence of the~parareal scheme. The reasons for its suitability remains the~same as for \textit{Cutoff Method}, even better convergence is expected.
        
    \paragraph{}
    By analyses of several options for $\mathcal{G}$, we would like to study, implement, and experimentally evaluate especially the~approaches based on cheaper methods for calculation of long-range interactions.
    
    \subsection{Complexity and ability to run in parallel}
Let's assume we are running a~simulation with the~parareal multilevel summation method with $T_W$ time points in computational window, $T_{total} = W T_W$ time points in whole simulation and $K$ iterations of convergence. For the~computation of $\lambda_n^{K-1}$ for all $n=1..T_W$, we need $T_W (K+1)-1$ calculations of function $\mathcal{G}$, $TK$ calculations of function $\mathcal{F}$ and $TK$ subtractions to get~$\Delta$. Assume we can calculate the~function $\mathcal{G}$ with $Q_{\mathcal{G}}$ floating point operations on $P_{\mathcal{G}}$ processors in time $R_{\mathcal{G}}$ and the~function $\mathcal{F}$ with $Q_{\mathcal{F}}$ floating point operations on $P_{\mathcal{F}}$ processors in time $R_{\mathcal{F}}$. If the~number of processors depends only on spatial decomposition of calculation, then $P_{\mathcal{G}} \approx P_{\mathcal{F}}$ as we are simulating exactly the~same system. For simplicity, let's assume that 
the overhead of calculation caused by synchronization and awaiting communication should be similar both for $\mathcal{G}$ and $\mathcal{F}$. Therefore, we can assume the ratio of two functions to be
\begin{equation}
\frac{R_{\mathcal{F}}}{R_{\mathcal{G}}} \approx \frac{Q_{\mathcal{F}}}{Q_{\mathcal{G}}}  = Q_{\mathcal{F}/\mathcal{G}}
\end{equation}

The cost of the~calculation heavily depends on the~distribution of tasks. In this context, a~task represents the~calculation of the~function $\mathcal{G}$ or $\mathcal{F}$ on given input data $\lambda$ and $\Delta$. The distribution of tasks assigns the~tasks to different processors. In preliminary analysis we came up with two distribution plans. Although they differ in number of used processors, the~time speed-up compared to the~sequential algorithm is directly proportional to $Q_{\mathcal{F}/\mathcal{G}}$ in both.

\paragraph{Calculation distribution plan 1}
The simplest plan assigns to computation unit P1\footnote{In these distribution plans, we consider a~computation unit to have $P_{\mathcal{G}}$ or $P_{\mathcal{F}}$ processors.} calculations of $\mathcal{G}$ for all time points in computational window whereas $T_W = Q_{\mathcal{F}/\mathcal{G}}$. That should take $Q_{\mathcal{F}/\mathcal{G}}R_{\mathcal{G}} = R_{\mathcal{F}}$ time. After that, $Q_{\mathcal{F}/\mathcal{G}}$ units calculate in parallel $\mathcal{F}$ for all points in the~computational window. After that, unit P1 calculates $\mathcal{G}$ for subsequent time points in the~following computational window and continues analogously. We need only $Q_{\mathcal{F}/\mathcal{G}}$ units. The speed-up is
\begin{equation}
\frac{T_{total}R_{\mathcal{F}}}{2\frac{T_{total}}{Q_{\mathcal{F}/\mathcal{G}}}R_{\mathcal{F}}} = \frac{Q_{\mathcal{F}/\mathcal{G}}}{2}
\end{equation}

\paragraph{Calculation distribution plan 2}
Figure \ref{fig:plan} depicts a~more pipelined plan. It assigns the~task of $\mathcal{F}$ calculation to the~available (or new) unit as soon as it has the~prerequisite---result of $\mathcal{G}$. We need $K$ units for $\mathcal{G}$ tasks and $KQ_{\mathcal{F}/\mathcal{G}}$ units for parallel calculation of $\mathcal{F}$. Time speed-up is
\begin{equation}
 \frac{TR_{\mathcal{F}}}{(\frac{T}{Q_{\mathcal{F}/\mathcal{G}}}+K)R_{\mathcal{F}}} = \frac{Q_{\mathcal{F}/\mathcal{G}}}{1+\frac{K}{TQ_{\mathcal{F}/\mathcal{G}}}}
\end{equation}
The value of $\frac{K}{TQ_{\mathcal{F}/\mathcal{G}}}$ would be close to 0 in case of the~method with high speed-up (proportional to $Q_{\mathcal{F}/\mathcal{G}}$) and good convergence (small $K$) applied to long simulation time (high $T$).
\begin{figure}[tbh]
\caption{Calculation distribution plan 2.}
\label{fig:plan}
\begin{center}
\includegraphics[width=\textwidth]{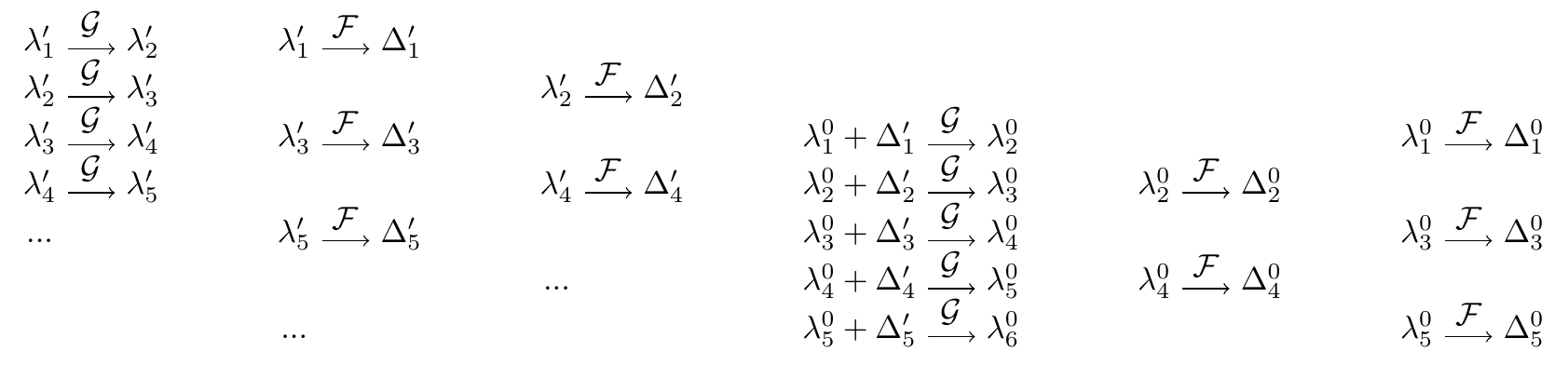}
\end{center}
\end{figure}

\paragraph{Example}
Let's assume $\mathcal{G}$ is simple cutoff method with cutoff 12\,\AA; $\mathcal{F}$ is MSM with $a=12$\,\AA~and $h=2$\,\AA. Then for $h^* = 1$\,\AA, $m=2$ (for $C^2$ Taylor smoothing function), $p=3$ (Hermite cubic interpolant)
\begin{equation}
\begin{gathered}
Q_{\mathcal{G}} = 2311N\\
Q_{\mathcal{F}} = 136813N\\
Q_{\mathcal{F}/\mathcal{G}} \approx 60\\
\end{gathered}
\end{equation}

Two calculation distribution plans presented above have time speed-up proportional to $Q_{\mathcal{F}/\mathcal{G}}$. With appropriately chosen functions we believe we can achieve the~speed-up by an~order of magnitude. Further distribution plans will be analyzed. 
\comment{
\paragraph{}

The calculation can accelerate if we can:
\begin{itemize}
\item Make $\mathcal{G}$ cheaper by using the~function $\mathcal{G}$ with shorter cutoff $a$ and changing grid spacing $h$. This would increase $Q_{\mathcal{F}/\mathcal{G}}$, therefore speed it up.   
\item Make $\mathcal{F}$ cheaper by calculating error $\Delta = \mathcal{F} - \mathcal{G}$ directly and approximate it instead of full calculation of $\mathcal{F}$. This would decrease $R_{\mathcal{F}}$ thus reduce the~calculation time proportionally.
\end{itemize}
{Dopisat tie veci ohladom odhadu chyby a~adaptivneho riadenia.}}

\subsection{Convergence}
The convergence of the~parareal MSM method, i.e. how high the~number of iterations is needed to satisfy the~acceptance criteria, influences the~speed-up and ability to run in parallel in indirectly proportional manner. In case of low number of iterations needed, the~width of the~computational window can grow and vice versa. The convergence depends mainly on the~accuracy and stability of $\mathcal{G}$. Both cutoff method and Wolf summation method are considered rather accurate \cite{Beck2005, Wolf1999}, the~simulations do not blow up just because of their errors. Although they do not achieve the~accuracy of more sophisticated methods, their cost in number of floating point operations is the~lowest. 

The error $\Delta$, the~difference between $\mathcal{F}$ and $\mathcal{G}$, decreases with increasing~cutoff. It represents the~change in positions of atoms calculated by $\mathcal{F}$ and positions calculated by $\mathcal{G}$. With increasing number of iterations, the~error grows exponentially, however, that does not necessarily mean high divergence. Two different sets of atom positions can in fact represent the~same system \cite{Yu2006, Srinivasan2005} if their radial distribution functions coincide. On the other hand, the similarity of atom positions from early time points does not necessarily mean good convergence during the whole simulation time. Therefore, the~actual convergence of the~method has to be evaluated by computer experiment with a prototype implementation. 

\subsection{Future work}
The presented proposal of the~parareal MSM method as parallel-in-time computation of MD simulations offers several opportunities for future work.

Combined with approach of Yu et al. \cite{Yu2006}, data from previously done coarse-grained simulation could serve as an~assessment of changes in system. The difference between simulations in vacuum and solvated systems would probably cause too many failures of their prediction algorithm, however, the~data can serve as high-level estimation. 

According to our analyses, we anticipate the~speed-up by an~order of magnitude. Further acceleration can be achieved through combination of spatial and temporal decomposition, already mentioned in \cite{Maday2005}. This concept has been successfully implemented for gravitational N-body problem in \cite{Speck2012}. 

Also, we do not neglect the~acceleration that GPUs can provide. The parareal scheme in the~second distribution plan nicely flows without major load inbalances if the number of iterations is fixed. The ratio between number of floating point operations and required memory transfers makes the~implementation on GPU platform worth trying. Moreover, GPUs implement the evaluation of inverse square root (for Euclidean distance between atoms) in hardware so the number of floating point operations needed is lower. That multiplies the ratio $Q_{\mathcal{F}/\mathcal{G}}$ and makes the speedup (compared with GPU calculation sequential-in-time) by two orders of magnitude possible. The number of supposedly saturated GPUs is directly proportional to the~width of the~computational window.

The parareal scheme combined with MD algorithm could conceal an~interesting possibility. If we are able to roughly approximate $\Delta$ without full evaluation of expensive $\mathcal{F}$, we can save the~calculation when the~correction is not necessary. Then, an~adaptive control of the~simulation could quickly calculate results in phases when the~system just moves around the minimum a~little bit. The interesting part with major changes would be evaluated by the~accurate function step-by-step. 
\comment{
{kam by sme to chceli posunut, pripadne ake su dalsie smery vyskumu}
\begin{itemize}
 \item use the~data from previous coarse-grained simulations -> maybe combination with \cite{Yu2006}
 \item synergic effect when combined with spatial parallelism, mentioned in \cite{Maday2005}, applied for gravitational N-body in \cite{Speck2012}
 \item the~assessment of the~error
 \item adaptive control
 \item GPU acceleration
 \item data compression for output files
\end{itemize}}

\subsection{Evaluation}
\comment{\begin{itemize}
 \item what will be evaluated - accuracy, strong scalability, weak scalability, speed, speed-up to single-cpu performance, correspondence with experimental results
 \item criteria - RMSD, relative error, ns/day
 \item 
\end{itemize}
}
We will evaluate several aspects of the~proposed methods. As many of them require experimental evaluation, we will develop a~prototype implementation. With data acquired by computer experiments we will compare accuracy and efficient parallelization with other parallel programs. 

First, we will examine the~method by feasible simulations of medium size and length at computational resources provided by CERIT-SC and Metacentrum NGI, Czech Republic. We will compare the~accuracy of the~results acquired by the~parareal MSM and by sequential-in-time MSM. The accuracy of MSM has been extensively analyzed and compared with other methods in \cite{Hardy2006}. Our comparison should uncover possible issues introduced by time parallelism. Strong scalability analysis can show its potential even with a~few hundreds of cores, if the~ratio between the~number of atoms and number of cores gets close to 0 (as achieved by Copernicus). Also, we will examine the~speed-up from classical MSM to parareal MSM to unfold the~contribution of time parallelism. Another measure of parallel MD code is the~speed in $ns/day$, i.e. how many ns of simulation time we can compute in 24 hours. However, that value depends on the~simulated system and the~parameters of the~infrastructure, so it needs to be interpreted in that 
context.

If our approach shows promising results on medium-sized simulations, we will proceed with more elaborate and optimized implementation. The large and long simulation of the~virus passing the~cell's membrane, infeasible with current methods, would require high number of computational resources. PRACE Research Infrastructure and ScalaLife project offer access to supercomputers in Europe, e.g. to half a~million cores in J\"{u}lich SuperComputing Centre or the~supercomputer with expected performance 1\,PFLOP/s that is being built in Ostrava, Czech Republic, by IT for Innovations, National Supercomputing Centre. With such large and long simulation run, we can compare the~weak and strong scalability of our approach with the~methods mentioned in subsection \ref{sec:parallel}. Moreover, as Pavel Plevka, CEITEC, will study this process by experimental means, we plan to compare the~results of our simulation to experimentally acquired ones. 
\end{section}
\newpage
\begin{section}{Aim of the~Work}
\comment{\begin{itemize}
 \item state the~question again
 \item what will be known afterwards that is not known now - if time parallelism is potential approach to partially solve the~sampling problem
 \item what will be created - the~implementation
 \item what are the~hypothesis to be investigated - if time parallelism will converge and give meaningful results, if it will speed up the~calculation, if it will increase strong scalability, if we can utilize large number of processors, if we can utilize GPUs
 \item what methods will be applied - prototype implementation, evaluation as descried earlier
\end{itemize}}

Proposed thesis aims to research how we can simulate large molecular dynamics systems for long simulation times. The acceleration of parallel implementations hits the~strong scalability limit. We want to increase the~level of parallelism by introducing the~concept of parallel-in-time computation to molecular dynamics.

Published solutions with time parallelism have both promising \cite{pronk11} and disappointing results \cite{Baffico2002}. The combination of the~parareal method and MD code has not achieved high speed-up or reasonable convergence, however, we suspect that inappropriately chosen coarse function cause that. We will further study and develop more suitable coarse functions. After this thesis, it will be known whether the~fine-grained time parallelism without \textit{a priori} knowledge increases the~strong scalability of MD algorithm. 

We will investigate several hypotheses through extensive study and evaluation of experiments with prototype implementation. We will analyze how different coarse functions affect the~convergence and speed-up of the~calculation and if the~method gives results comparable with sequential-in-time methods in accuracy. Our concern will focus on the~strong scalability---how many cores we can saturate with decreasing time to result and how much we can reduce the~ratio between the~size of the~system and the~number of resources. 

\subsection{Expected Results}
\comment{\begin{itemize}
\item contribution to the~area
\item increased strong scalability
 \item the~implementation of MSM+parareal algorithm
 \item evaluation of algorithm with toy systems (only water) and with the~system simulating the~virus passing a~cell's membrane provided by dr. Vácha, compared to experiments of Mgr. Pavel Plevka, Ph.D.
 \item scientific articles published in related conferences and journals
 \item the~dissertation thesis describing the~analysis, proposed solution, method's details, and evaluation along with technical documentation
\end{itemize}}

We expect to contribute to both computational chemistry and computer science. If successful, the~time parallelism could, to some extent, solve the~sampling problem, and serve as inspiration for other computational chemistry tools to incorporate it into their calculations. Research in high performance computing is more and more interested in algorithms that are able to utilize large computational resources.  From the~point of decomposition algorithms for parallel and distributed computing, an~application of the~combined spatial and temporal decomposition shows perspective for other simulations. 

We will implement a~prototype algorithm to experimentally evaluate the~accuracy, convergence, speed-up, and strong scalability on systems with feasible size and simulation time. If the~prototype algorithm shows high speed-up and scaling we want to apply for computational time of a~supercomputer. A~simulation of the~virus passing the~cell's membrane with possibly hundreds of millions atoms for at least 50\,$\mu s$ would verify properties of the method at large scale.

We plan to publish the~method and results in related conferences and scientific journals, such as:
\pagebreak

\textbf{Conferences}
\begin{itemize} 
 \item IEEE International Conference on High Performance Computing and Simulation 
 \item ACM/IEEE/SCS Workshop on Parallel and Distributed Simulation
 \item The International ACM Symposium on High-Performance Parallel and Distributed Computing
 \item International Conference on High Performance Computing \& Simulation
 \item International Conference on High Performance Computing
 \item Supercomputing Conference
 \item Summer Computer Simulation Conference
 \end{itemize}
 \textbf{Scientific Journals}
 \begin{itemize}
 \item Journal of Computational Chemistry
 \item Journal of Computational Physics
 \item Journal of Chemoinformatics
 \item Journal of Chemical Information and Modeling
 \item Journal of Chemical Theory and Computation
 \item Journal of Chemical Physics
 \item Computer Physics Communications
 \item Parallel Algorithms and Applications
 \item SIAM Journal on Scientific Computing
 \item Parallel Computing
 \item Domain Decomposition Methods in Science and Engineering
 \item Journal of Parallel and Distributed Computing 
\end{itemize}

The text of PhD. thesis will, as common, analyze the~field, state an~interesting and difficult problem, propose a~novel solution and evaluate it. 

\subsection{Schedule}
We schedule future progress with the following plan. 
\begin{itemize}
 \item 01/2014 - 05/2014: implementation of the prototype MD code with MSM for evaluation of long-range interactions
 \item 06/2014 - 07/2014: evaluation of the convergence of the parareal scheme, conference article writing
 \item 06/2014 - 11/2014: implementation of the parareal scheme with MSM as the fine function and cutoff/Wolf summation method as coarse functions 
 \item 12/2014 - 01/2015: evaluation of the speed-up and strong scalability of the parareal scheme on feasible simulations, journal article writing
 \item 01/2015 - 06/2015: optimizations of the prototype 
 \item 06/2015 - 12/2015: implementation of GPU acceleration, conference article writing, applying for computing time for the simulation of the virus
 \item 01/2016 - 06/2016: thesis writing, the simulation run
 \item 06/2016: submission of the thesis
 \item 06/2016 - 12/2016: journal article writing, postprocessing data from the simulation run
 \item 12/2016: the thesis defense
\end{itemize}

\end{section}

\glsaddall
\printglossaries

\begingroup
\raggedright
\sloppy
\bibliography{librarybib}
\endgroup

\end{document}